\documentclass[preprint]{aastex61}

\received{November 7, 2017}
\revised{March 15, 2018 }
\accepted{March 27, 2018 }
\submitjournal{ApJ}

\shorttitle{solar flare reconnection}
\shortauthors{Forbes et al.}

\begin{document}

\title{Reconnection in the Post-Impulsive Phase of Solar Flares}

\correspondingauthor{Terry G. Forbes}
\email{terry.forbes@unh.edu}

\author[0000-0003-0538-3509]{Terry G. Forbes}
\affiliation{Institute for the Study of Earth, Oceans, and Space, University of New Hampshire, 8 College Road, Durham, NH 03824, USA}

\author[0000-0002-0494-2025]{Daniel B. Seaton}
\affiliation{Cooperative Institute for Research in Environmental Sciences, University of
Colorado,  Boulder, CO 80305, USA}
\affiliation{NOAA National Centers for Environmental Information,  Boulder, CO 80305, USA}
\affiliation{Solar-Terrestrial Center of Excellence, Royal Observatory of Belgium, Ringlaan-3-Av. Circulaire, B-1180 Brussels, Belgium}

\author[0000-0002-6903-6832]{Katharine K. Reeves}
\affiliation{Harvard-Smithsonian Center for Astrophysics, 60 Garden Street MS 58,
Cambridge, MA 02138, USA}



\begin{abstract}

Using a recently developed analytical procedure, we determine the rate of magnetic reconnection in the ``standard" model of eruptive solar flares.  During the late phase, the neutral line is located near the lower tip of the reconnection current sheet, and the upper region of the current sheet is bifurcated into a pair of Petschek-type shocks.  Despite the presence of these shocks, the reconnection rate remains slow if the resistivity is uniform and the flow is laminar.  Fast reconnection is achieved only if there is some additional mechanism that can shorten the length of the diffusion region at the neutral line.  Observations of plasma flows by the X-Ray Telescope (XRT) on {\it Hinode} imply that the diffusion region is in fact quite short.  Two possible mechanisms for reducing the length of the diffusion region are localized resistivity and MHD turbulence.

\end{abstract}




\section{Introduction}

Although most models of eruptive flares incorporate magnetic reconnection, they often do so in an ad hoc way.  For example, the analytical loss-of-equilibrium model of \citet{lin00} and \citet{reeves05} assumes that reconnection occurs at a neutral point located at the center of a post-eruption current sheet.  The model also assumes that the plasma flows into this current sheet at a constant Alfv\'{e}n Mach number whose value is treated as a free parameter.  Even in numerical models, a realistic prescription of the reconnection process is often lacking because of inadequate numerical resolution of the current sheets in which reconnection occurs \citep{matthaeus81}.  Recently, we developed an analytical theory that predicts the reconnection rate and the location of the neutral point in both symmetric \citep{forbes13} and asymmetric configurations \citep{baty14}.  Quantitative comparisons with two-dimensional, resistive MHD simulations show that the theory successfully predicts the reconnection rate and the location of the neutral point to an accuracy of 5 to 10\% as long as the simulation is carried out in the laminar regime \citep{baty14}.  Here we use this theory to replace the ad-hoc assumptions of \citet{lin00} and \citet{reeves05} with a prescription of the reconnection process that is physics based.

The analytic theory that we use predicts that Sweet-Parker reconnection \citep{parker57} occurs when the plasma resistivity is uniform and the magnetic field is symmetric \citep{forbes13}.  This kind of reconnection is too slow to account for the rapid energy release in flares unless the resistivity of the plasma in the corona is many orders of magnitude higher than expected \citep{priest02}.  However, if the resistivity is not uniform, or the field is not symmetric, then our theory predicts that Petschek-type reconnection \citep{petschek64} may occur.  Pairs of slow-mode shocks emanating outward from a diffusion region are a key feature of this kind of reconnection.  Whether the reconnection is fast or not depends upon the length of the diffusion region relative to the global scale of the erupting magnetic field.  In this paper we assume for simplicity that the resistivity is uniform.  Thus, any slow shocks that occur are due to the asymmetry of the magnetic field.  In the eruptive flare model that we consider the asymmetry is caused by the decrease of the coronal magnetic field with radial distance.  This decrease creates a vertical current sheet whose field is strong near the solar surface but weak at high altitude as shown in Figure 1.  This configuration is sometimes referred to as the "standard" model for the gradual phase of solar flares \citep{janvier14}.
\begin{figure}
\plotone{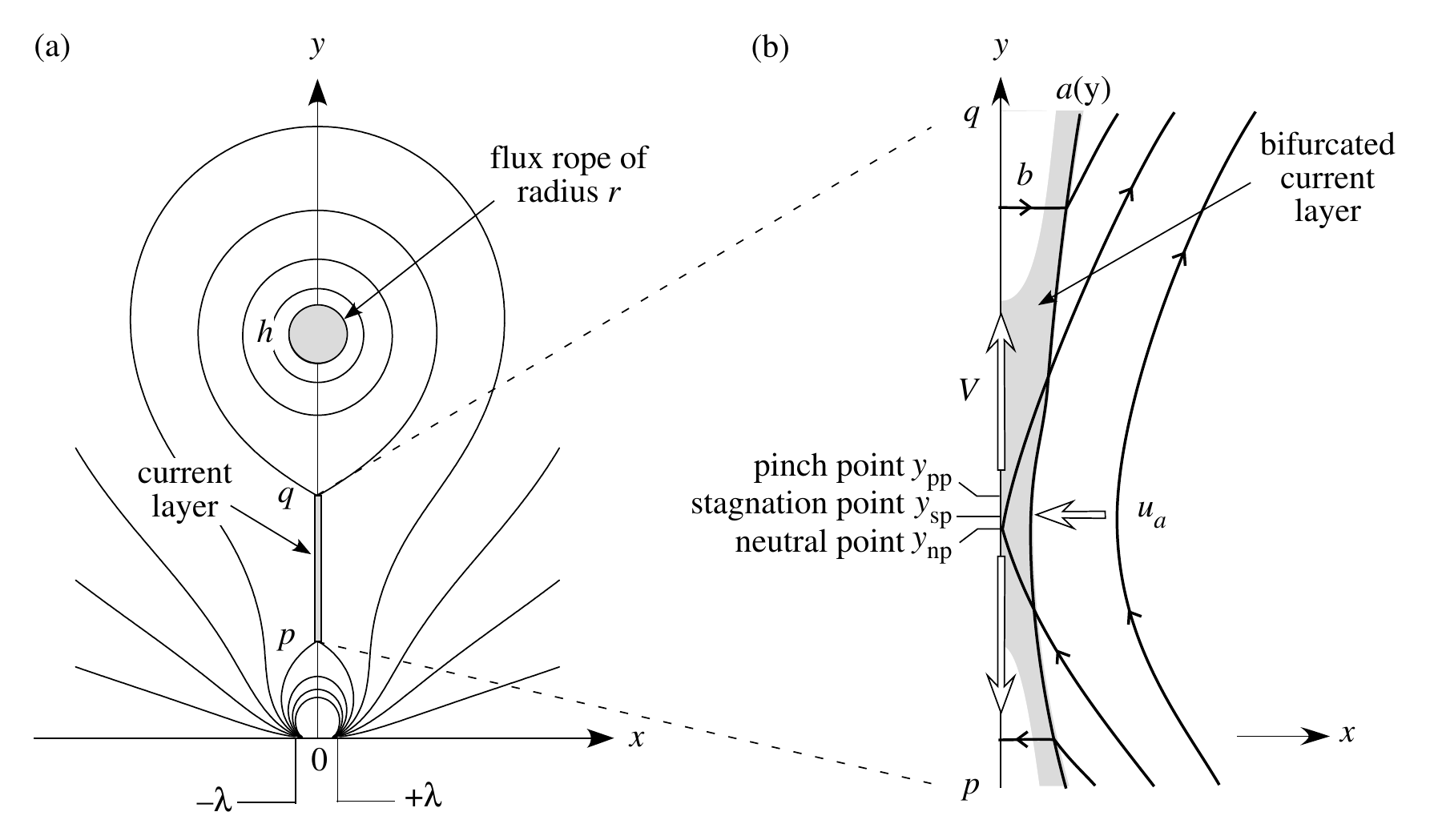}
\caption{Magnetic field configuration of the eruptive flare model of \citet{reeves05} with embedded current sheet.  The left diagram (a) shows a flux rope of radius $r$ centered at the height $h$.  The locations $q$, and $p$ correspond to the upper and lower tips of a current sheet located on the $y$-axis.  The field at the solar surface is represented by point sources located at $\pm \lambda$.  The right diagram (b) shows a close up of the bifurcated current sheet whose half thickness is $a(y)$.  The plasma flows into the sheet with the velocity $u_{a}(y)$ and out of the sheet with the velocity $V(y$).  The locations $y_{pp}$, $y_{sp}$, and $y_{np}$ correspond to the maximum tangential magnetic field (i.e. the pinch point where $\partial B_{ya}/ \partial y = 0$), the stagnation point ($V = 0$), and the neutral point ($b = 0$).}
\end{figure}
 The magnetic field, ${\bf B}$, in Figure 1 is prescribed by
 \begin{equation}
 B_{y} + iB_{x} = {2 i A_{0} \lambda (h^2 + \lambda^2) \sqrt{(z^2 + p^2) (z^2 + q^2)} \over \pi (z^2 - \lambda^2) (z^2 + h^2) \sqrt{(\lambda^2+p^2) (\lambda^2+q^2) }} \; {\mbox{\rm for}} \; |z-ih| \geq  r ,
 \end{equation}
where  $z = x + iy$.  Here $A_{0}$ is the magnitude of the vector potential at the origin, $\lambda$ is the half-distance between the field sources at $y = 0$, $h$ is the height of the flux rope, $p$ is the height of the lower tip of the current sheet, and $q$ is the height of the upper tip.   The formula for the corresponding vector potential, ${\bf A}(z)$ can be found in  \citet{reeves05}.  Inside the flux rope ($|z-ih| < r$) the field is prescribed by a force-free solution due to \citet{parker74}.  The flux rope current, $I$, is related to $h$, $p$, and $q$ by
\begin{equation}
I = {c \lambda A_{0} \over 2 \pi h} {\sqrt{(h^2-p^2) (h^2-q^2)} \over \sqrt{(\lambda^2 + p^2) (\lambda^2+q^2)}} ,
\end{equation}
where $c$ is the speed of light.  Expression (2) comes from the requirement that the magnetic field at the tips of the current sheet be zero \citep{lin00}.

The analytical theory we use also predicts the location of the magnetic neutral point within the current sheet.  In the absence of an imposed symmetry, predicting the location of the neutral point from theory is just as challenging as predicting the rate of reconnection.  Previously, \citet{reeves05} assumed that the neutral point was located in the center of the flare current sheet, but, as we will show, this assumption is not valid because of the asymmetry introduced by the decrease of the solar magnetic field with height.

In the next section, we present the analytical theory that we use to describe the reconnection process in the post-eruption current sheet.  Then in section 3, we apply this theory to the loss-of-equilibrium model previously considered by \citet{lin00} and \citet{reeves05}.  In section 4 we discuss the observational significance of our results, and then, in section 5 we present our conclusions.

\section{Reconnection Rate and Location}

To incorporate the physics of the reconnection process into the eruptive flare model we use a method that simplifies the reconnection problem by averaging the resistive-MHD equations over the reconnection current sheet \citep{forbes13,baty14}.  The idea of averaging the equations in this way was first considered by \citet{vasyliunas75} over 40 years ago for an incompressible plasma, and 10 years later by \citet{titov85a} for a compressible one (see also \citealt{somov92}).  Both of these previous studies obtained steady-state solutions for the field and flow within the current sheet, but it has only recently become evident that most of these solutions are structurally unstable and, therefore, unphysical \citep{forbes13}.  These unstable solutions contain an essential singularity at the stagnation point between the two reconnection-outflow jets.  However, in some circumstances solutions may exist that do not contain such a singularity.  These nonsingular solutions are structurally stable, and physically obtainable.  Typically what is required for the existence of such solutions is a spatial nonuniformity of some sort.  The nonuniformity may occur in the electrical resistivity of the plasma or in the external magnetic field outside the current sheet.  In the few cases where the analytical solutions have been compared with resistive MHD simulations, the discrepancies between the two range from 5\% to 14\% \citep{baty14}.

Although the general method for calculating the reconnection rate and location allows for a time-dependent magnetic field \citep{forbes13}, a time-dependent analysis is not needed if we restrict our attention to the post-impulsive phase of the eruption.  As shown in Appendix A, time-dependent effects near the neutral point are negligible a few Alfv\'{e}n time scales after the onset of the eruption.  The primary reason why the reconnection process becomes steady is that its rate and location are controlled by the geometry of the magnetic field just above the flare loops.  These loops change very slowly in time during the post-impulsive phase, so the reconnection process is quasi-steady during this period.

   For a quasi-steady configuration the flow velocity, $V$, averaged across the thickness of the reconnection current sheet satisfies the differential equation \citep{titov85a, titov85b, somov92, seaton09, baty14}:
\begin{equation}
 I_{B} {\partial V \over \partial y} + { V \over B_{a}(y)} = {B_{a}(y) \over 4 \pi \rho_{a} V} \left(1-I_{B} { \partial B_{a}(y) \over \partial y} {\rho_{a} \over \rho}\right) - {\alpha B_{a}^2(y) \over B_{asp}^3 I_{B} \sqrt{4 \pi \rho_{a}}} \; {\eta \over \eta_{sp}} \; {\rho \over \rho_{a}},
\end{equation}
where $y$ is the coordinate along the length of the current sheet, $\rho$ is the average density within the current sheet, $\rho_{a}$ is the ambient density outside the current sheet, $\eta$ is the magnetic diffusivity, and $\eta_{sp}$ is the diffusivity at the location $y_{sp}$ of the stagnation point of $V$.  The magnetic diffusivity, $\eta$, is related to the electrical resistivity, $\eta_{e}$, by $\eta = \eta_{e} c^2/4\pi$ \citep{priest14}.  In general $\eta$ may be a function of space, time, or any of the plasma variables.  Here we assume it is uniform, so $\eta / \eta_{sp} = 1$.  $B_{a}(y)$ is the exterior component of the magnetic field parallel to the current sheet and just outside it.  In other words the $y$-component of the magnetic field at the location $x = a$ in Figure 1.  $B_{asp}$ is the value of $B_{a}(y)$ at $y_{sp}$.  The functional form of $B_{a}(y)$ is initially determined using an external field model for an infinitely thin, static current sheet (e.g.  \citealt{green65, syrovatskii71}).  If needed, $B_{a}(y)$ can be iterated to produce a more accurate expression once a solution for $V(y)$ is obtained (see Appendix A in \citealt{forbes13}).

The parameter $\alpha$ is defined by
\begin{equation}
\alpha = {\eta_{sp} \sqrt{4 \pi \rho_{a}} \over M_{Asp}^2 B_{asp}} ,
\end{equation}
where $M_{Asp}$ is the Alfv\'{e}n Mach number of the inflowing plasma at $x = a$, $y = y_{sp}$, immediately upstream of the current sheet at the location of the stagnation point (cf. Figure 1).  The parameter $\alpha$ corresponds to the half-length of the diffusion region in the incompressible reconnection theory of Sweet and Parker \citep{parker57}.  Although our analysis here is compressible, $\alpha$ still provides a reasonable estimate of the length scale, so we will refer to it as the diffusion-region-length scale.  However, it should be kept in mind that the actual length of the diffusion region depends weakly on several parameters such as the plasma beta and the functional form of $B_{a}$.  The reconnection rate, $M_{Asp}$, is expressed in terms of $\alpha$ as
\begin{equation}
M_{Asp} = \sqrt{{\eta_{sp} \sqrt{4 \pi \rho_{a}} \over \alpha B_{asp}}} .
\end{equation}

The density, $\rho$, of the plasma in the current sheet is given by
\begin{equation}
\rho = {\gamma (B_{a}^2(y) + \beta B_{0}^2) / (\gamma - 1)\over \gamma \beta B_{0}^2 / (\gamma-1) - 4 \pi \rho_{a} V^2 + 2 J_{B} / I_{B}} \; \rho_{a} ,
\end{equation}
where $\gamma$ is the ratio of specific heats (i.e. 5/3), $\beta = 8 \pi p_{a} / B_{0}^2$ is the upstream plasma beta of the inflow region, and $B_{0} = A_{0} / \lambda$, is the average of the vertical magnetic field at $y = 0$ from $x = 0$ to $\lambda$.  Finally, the functions $I_{B}$ and $J_{B}$ are defined as
\begin{eqnarray}
I_{B}(y, y_{sp}) =  \int^y_{y_{sp}} {dy' \over B_{a}(y')}, \; \; {\mbox{\rm  and }} \; J_{B}(y, y_{sp}) = \int^y_{y_{sp}} B_{a}(y')dy' . \nonumber
\end{eqnarray}

The corresponding solutions for the current-sheet thickness, $a(y)$, and the transverse field, $b(y)$, within the current sheet are given by the auxiliary equations:
\begin{equation}
a(y) = {\rho_{a} \over \rho V} {M_{Asp} B_{asp}^2 \over \sqrt{4 \pi \rho_{a}}} I_{B} ,
\end{equation}
and
\begin{equation}
b(y) = {M_{Asp} B_{asp}^2 \over V \sqrt{4 \pi \rho_{a}}} - {B_{a}(y) \; \eta \over V a} .
\end{equation}
The current density averaged across the sheet is $j = c B_{a}(y) / (4 \pi a)$.

Several assumptions are made in obtaining Equation~(3) as follows:

\begin{enumerate}

\item The inflow Alfv\'{e}n Mach number, $M_{Asp}$, is assumed to be much less than one.  This assumption allows the MHD equations to be expanded in terms of the small parameter $M_{Asp}$ (see \citealt{erkaev02, forbes13}).  Quantities like $a$, $u_{a}$, and $b$ are then of order $M_{Asp}$, and terms that are of second order or higher are neglected.

\item The external flow, $V_{a}$, parallel to the current sheet, is assumed to be negligible (i.e. of order $M_{Asp}$ or smaller).  This particular assumption is valid for Sweet-Parker and Petschek reconnection, but not necessarily for other types of reconnection such as flux pile up \citep{priest86}.

\item The quantities $\rho$, $V$, and $b$ are assumed to be nearly uniform in $x$ within the current sheet.  This assumption allows averages of a product, like $\rho V$, to be expressed as a product of the individual averages of $\rho$ and $V$.

\item The variation of quantities in the direction of the outflow is assumed to be relatively smooth so that
gradient operator, $\partial / \partial y$, is of zero order in the expansion parameter, $M_{Asp}$.

\item The parallel magnetic field, $B_{y}$, within the current sheet is assumed to be of order $M_{Asp}$ or smaller.

\item The flow is assumed to be laminar and stable.  As we will discuss in Section 4, this assumption holds as long as the Lundquist number is less than $\approx 10^4$.

\item The energy equation used to derive Equation~(3) does not include losses due to thermal conduction or radiation.

\end{enumerate}

Although the present analysis neglects thermal conduction, we expect it to be important within the current sheet.  Thermal conduction drains thermal energy out of the sheet, which both cools and slows the plasma \citep{somov00, seaton09}.   A numerical simulation by \citet{yokoyama96} found that the reconnection rate increases only by about 20\% when thermal conduction is added.  The lack of any dramatic change in the reconnection rate may be due to the fact that a nonuniform resistivity of fixed length was used to control the length of the diffusion region in their simulation.  We would expect that if a temperature-dependent resistivity model had been used instead, then thermal conduction would have had a major effect on the rate of reconnection.

A comparison of the analytical solutions with resistive-MHD simulations shows that one of the larger sources of error is due to assumption 3.  In low beta plasmas there are density variations across the width of the current sheet that generate errors on the order of 5\% to 10\% in the reconnection rate and on the order of 3\% in the location of the stagnation point \citep{baty14}.  A detailed derivation of Equation~(3), as well as additional discussion of the assumptions used to obtain it, can be found in \citet{seaton09}, \citet{forbes13}, and \citet{baty14}.

Equation~(3) together with Equation~(6) constitutes a first order differential equation for the outflow velocity, $V$.  It is similar to the MHD nozzle equation that is often used to model astrophysical jets, except that it includes resistivity.  In the limit that $\beta \rightarrow \infty$, the equation reduces to the one first derived by \citet{vasyliunas75} for an incompressible plasma.  The constant of integration associated with Equation~(3) is determined by the requirement that the solution contain a stagnation point (see \citealt{forbes13}).  Once this condition is imposed, the integration constant is fixed, and Equation~(3) yields a solution for $V$ in terms of the unknown constants $y_{sp}$ and $\alpha$.  Solutions of this type can be found in \citet{titov85b}, \citet{somov87}, and \citet{somov92}.

What has not been realized until quite recently is that most solutions of Equation~(3) are unphysical because they contain an essential singularity at the stagnation point.  In the time-dependent system, the singular solutions are structurally unstable and rapidly collapse \citep{forbes13}.  Most solutions are unstable, but stable solutions (i.e. nonsingular ones) may exist for special values of $y_{sp}$ and $\alpha$.  Nonsingular solutions typically occur when there is a nonuniformity of some sort in the system, for example, a nonuniform resistivity or a nonuniform $B_{a}$.  The nonuniformity must be such that it generates a transverse field component, $b$.  When the length scale of the nonuniformity is less than the length of current sheet, a Petschek-type configuration, with slow-mode shocks, appears \citep{forbes13}.

Because the \citet{reeves05} model assumes the gas pressure in the background corona is negligible, we set $\beta = 0$ in Equation~(6).  (Recall that $\beta$ is the ratio of the gas to magnetic pressure in the inflow region upstream of the current sheet.)  Also setting $\gamma = 5/3$, we obtain
\begin{equation}
\rho = {5 B_{a}^2(y) \over 4 J_{B}/I_{B} - 8 \pi \rho_{a} V^2}\; \rho_{a} ,
\end{equation}
for the density within the current sheet.  To separate the nonsingular solutions from the singular ones, we expand $V$ and $B_{a}$ in power series centered on the stagnation point, $y_{sp}$:
\begin{eqnarray}
V(y) =  \sum^\infty_{n=1} {V_{n} (y-y_{sp})^n}, \; \; {\mbox{\rm  and }} \; B_{a}(y) = \sum^\infty_{n=0} B_{an}(y-y_{sp})^n . \nonumber
\end{eqnarray}
Substitution of these series into Equations (3) to (6) with $\beta = 0$ yields the first three terms:
\begin{equation}
V_{1} = 4 B_{a0}/(5 \alpha \sqrt{4 \pi \rho_{a}}) = 4B_{asp}/(5 \alpha \sqrt{4 \pi \rho_{a}}) ,
\end{equation}
\begin{equation}
V_{2} = -66 B_{a1}/(25 \alpha \sqrt{4 \pi \rho_{a}}) ,
\end{equation}
\begin{equation}
V_{3} = (-96 B_{a0}^2 + 462 \alpha^2 B_{a1}^2 - 256 \alpha^2 B_{a0} B_{a2})/(75 B_{a0} \alpha^3 \sqrt{4 \pi \rho_{a}}) .
\end{equation}
Requiring the series for $V$ to converge eliminates the singular solutions.  If the series converges, then $V$ is analytic at $y_{sp}$, and it can be approximated by a partial sum consisting of the first few terms of the series.

A necessary condition for such convergence is that the coefficients for $V$ tend to zero as $n$ tends to infinity.  That is
\begin{equation}
\lim_{n \to \infty} V_{n} = 0 .
\end{equation}
By contrast, singular solutions have coefficients that tend to infinity as $n \rightarrow \infty$ (see Appendix B of \citealt{forbes13}).  If the series converges, the values of $\alpha$ and $y_{sp}$ can be approximately determined by imposing the conditions that $V_{2m+1} = 0$ and $V_{2m} = 0$ where $m \geq 1$. The first condition is for the odd terms and the second for the even ones.  In configurations where the exterior field model is symmetric (e.g. the \citet{green65} and \citet{syrovatskii71} models), only the first condition is needed since all the even terms in the series for $V$ will be zero.  The approximate values become increasingly more accurate as $m$ increases.  The lowest order approximation for the location of the stagnation point and the reconnection rate is obtained by setting $V_{2}$ and $V_{3}$ to zero.  The equation $V_{2} = 0$ immediately leads to
\begin{figure}
\plotone{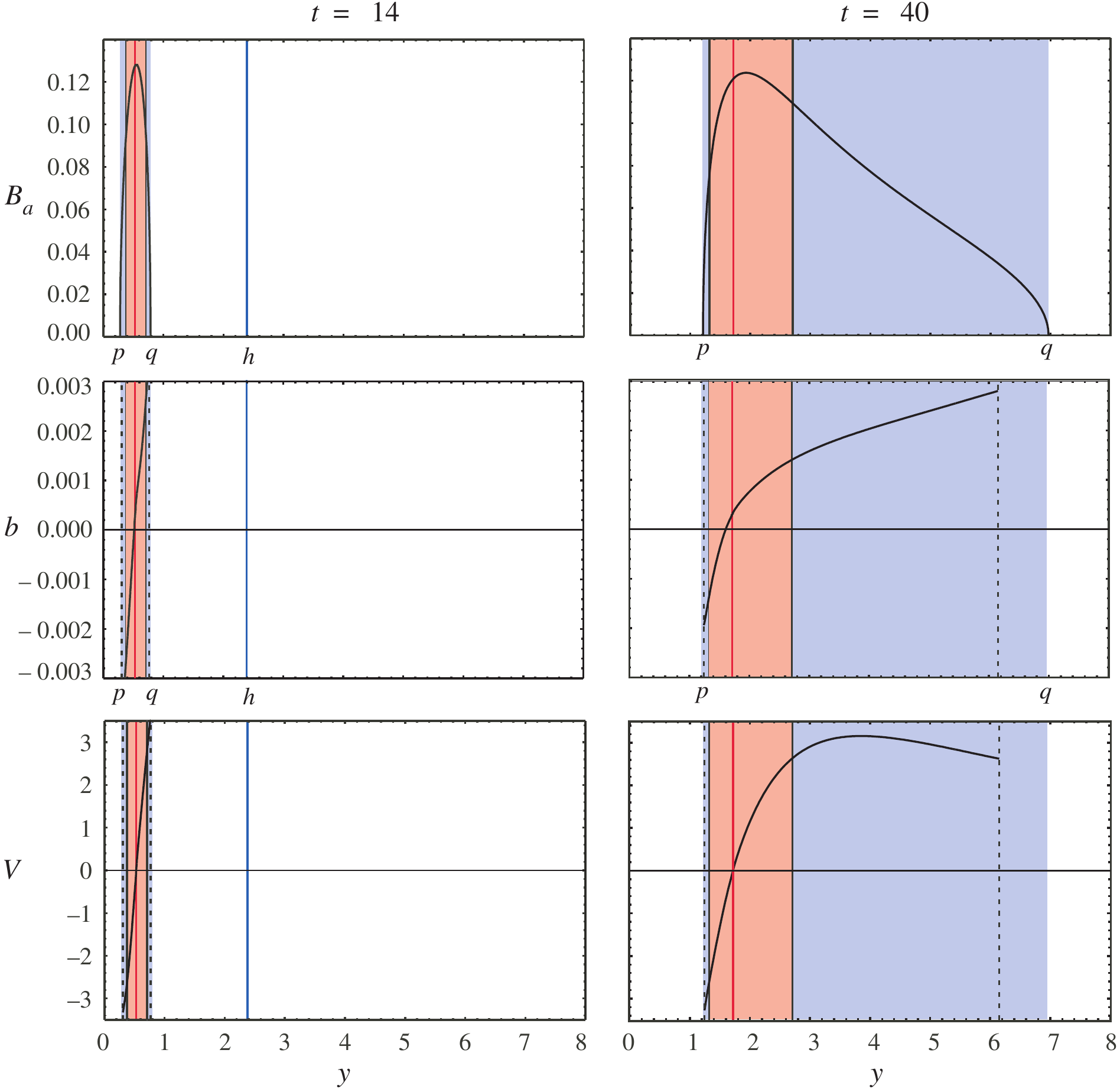}
\caption{Exterior longitudinal magnetic field component $B_{a}$, interior transverse magnetic field component, $b$, and reconnection outflow velocity $V$, as functions of distance $y$ along the current sheet at two different times, $t$.  Salmon colored shading indicates the diffusion region where the diffusive electric field, $\eta j$, is greater than the advective electric field, $V b$, while blue shading indicates the advective region where the reverse is true.  The red vertical line shows the location of the stagnation point, and the blue vertical line shows the location of the flux rope's center.  (There is no blue line at $t = 40$, because by this time the flux rope's center has reached a height of $y = 16.6$.)  Dashed vertical lines mark the locations where the expansion procedure used to obtain the solutions breaks down.  Lengths are normalized to $\lambda_{0}$, $B_{a}$ and $b$ are normalized to $A_{0} / \lambda_{0}$, and $V$ is normalized to $V_{0}$ (see Section 3).}
\end{figure}
\begin{eqnarray}
B_{a1} = 0 \nonumber ,
\end{eqnarray}
or equivalently,
\begin{eqnarray}
 \frac{dB_{a}}{dy}\Bigr|_{{y_{sp}}} = 0,
\end{eqnarray}
which means that $y_{sp}$ is approximately located at the pinch point, $y_{pp}$.  The pinch point is defined as the location where $B_{a}(y)$ has a maximum, that is where $dB_{a}(y)/dy = 0$ and $d^2B_{a}(y)/dy^2 < 0$.  At the pinch point the external magnetic field lines bow inward as shown in Figure 1b.  Thus, to lowest approximation, the stagnation point of the flow is located close to where one expects a neutral point to form, that is:
\begin{equation}
y_{sp} \approx y_{pp} .
\end{equation}
Consequently, setting $V_{3}$ to zero yields
\begin{eqnarray}
 \alpha \approx  \sqrt{-3B_{a0}/(8 B_{a2})} \nonumber,
\end{eqnarray}
or, in terms of derivatives
\begin{equation}
\alpha \approx \sqrt{-3B_{app}/(4 B_{app}'')} ,
\end{equation}
where $B_{app}''$ is the second derivative of $B_{a}$ evaluated at $y_{pp}$.  Since $B_{app} > 0$, we see that a stable solution exists only if $B_{app}'' < 0$.  This condition is always satisfied for the flare model current sheet.  Furthermore, we see that to lowest order the scale-length associated with the second derivative of $B_{a}$ at $y_{pp}$ determines the size of $\alpha$.

From Equation~(1) the field, $B_{a}$, immediately exterior to the positive side ($x > 0$) of the current sheet in the Reeves \& Forbes flare model is
\begin{equation}
 B_{a}(y) = \lim_{x \to 0} B_{y}(x,y) = {2 A_{0} \lambda (h^2 + \lambda^2) \sqrt{(y^2 - p^2) (q^2 - y^2)} \over \pi (y^2 + \lambda^2) (h^2 - y^2) \sqrt{(\lambda^2+p^2) (\lambda^2+q^2) }}  ,
\end{equation}
where  $p < y <  q$.  We now use this expression to evaluate the coefficients in Equations (11) and (12) for the two different cases shown in Figure 2.  The first case ($t = 14$) corresponds to a time early in the evolution of the flare model, and the second corresponds to a later time ($t = 40$).  For both cases the constant $\lambda$ is $0.9695 \lambda_{0}$ where $\lambda_{0}$ is the length scale used to normalize quantities in the flare model (see Section 3).  For $t = 14$, $p = 0.280 \lambda_{0}$, $q = 0.788 \lambda_{0}$, $h = 2.461 \lambda_{0}$.  While for $t = 40$, $p = 1.213 \lambda_{0}$, $q = 6.966 \lambda_{0}$, $h = 16.612 \lambda_{0}$.  (The time is normalized with respect to the time scale, $t_{0}$, used in the flare model discussed in Section 3).  The top panels of Figure 2 show $B_{a}$ as function of $y$ for the two sets of parameters.  At the early time the length, $q - p$, of the current sheet is shorter than the distance $2 \lambda_{0}$ between the photospheric source regions of the field, but at the later times it is significantly greater than this distance.
\begin{figure}
\epsscale{0.65}
\plotone{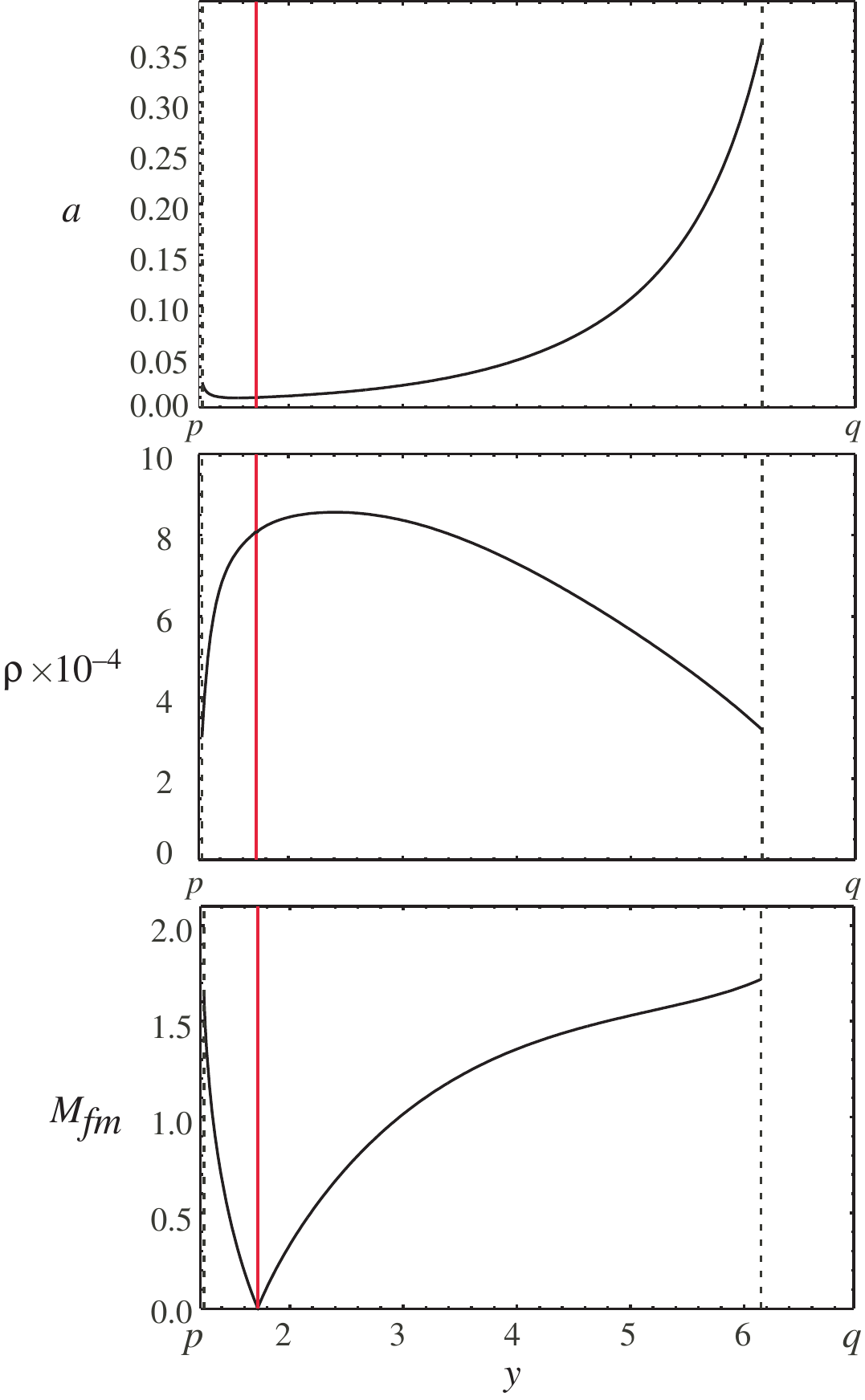}
\caption{Current sheet thickness, $a$, density, $\rho$, and outflow fast-mode Mach number $M_{fm}$. as functions of distance $y$ along the current sheet at $t = 40$.  The red vertical line shows the location of the stagnation point, while the dashed vertical lines mark the locations where the expansion procedure used to obtain the solutions breakdown.  Lengths are normalized to $\lambda_{0}$, and the density is normalized to $\rho_{0}$ (see Section 3).}
\end{figure}

   Table 1 shows the values of $y_{sp}$ and $\alpha$ that are obtained for different levels of approximation.  The top line of values are derived from Equations (14) and (16).  Subsequent values are obtained by setting the odd and even pairs of high-order coefficients to zero in the series expansion of $V$.  For the early time ($t = 14$) the values of $y_{sp}$ and $\alpha$ rapidly converge and are accurate to five significant figures when $n = 8$.  However, for the late time ($t = 40$) the values of $y_{sp}$ and $\alpha$ converge more slowly, reaching an accuracy of five significant figures only when $n = 13$.  The slower rate of convergence is due to the greater asymmetry of the magnetic field in the current sheet.

   Although the values of $y_{sp}$ and $\alpha$ in Table 1 are given to 5 significant figures, this does not mean that we have determined the reconnection rate and location to this degree of accuracy.  The one-dimensional nozzle equations are highly idealized, and they are unlikely to be accurate to more than 5\% \citep{schreier82,baty14}.  Therefore, we use a level of approximation that is consistent with the overall accuracy of the equations, namely the values obtained using $V_{3}$ \& $V_{4} = 0$.  Using these values we obtain the velocity curves shown in Figure 2.  The top two panels of Figure 3 show the solution for thickness, $a$, and the density, $\rho$, of the current sheet for $t = 40$.  These curves do not extend all the way to $p$ and $q$, because the assumption that the inflow Alfv\'{e}n Mach number is small starts to break down as one approaches the nulls of $B_{a}$ at $p$ and $q$.  The criterion used to define the location at which the expansion breaks down is $|da/dy| = 1/2$.  If the slope of $a(y)$ becomes too steep, then the assumption that variations parallel to the current sheet are small compared to those across it, no longer holds.

\begin{deluxetable}{rcccc}
\tablecaption{Successive Approximations for Stagnation-Point Height, $y_{sp}$, and Diffusion Region Length-Scale, $\alpha$}
\tablecolumns{5}
\tablewidth{0pt}
\tablehead{
\colhead{Level of} &
\colhead{$y_{sp}$\tablenotemark{*}} &
\colhead{$\alpha$\tablenotemark{*}} &
\colhead{$y_{sp}$\tablenotemark{*}} &
\colhead{$\alpha$\tablenotemark{*}} \\
\colhead{Approximation} &
\colhead{$(t = 14)$} &
\colhead{$(t = 14)$} &
\colhead{$(t = 40)$} &
\colhead{$(t = 40)$}}
\startdata
$V_{2} \; \& \; V_{3} = 0$ & 0.55204 & 0.20909 & 1.93045 & 0.97722 \\
$V_{3} \; \& \; V_{4} = 0$ & 0.55614 & 0.20841 & 1.76995 & 0.70734 \\
$V_{4} \; \& \; V_{5} = 0$ & 0.55562 & 0.20047 & 1.76331 & 0.70938 \\
$V_{5} \; \& \; V_{6} = 0$ & 0.55572 & 0.20045 & 1.76149 & 0.70296 \\
$V_{6} \; \& \; V_{7} = 0$ & 0.55573 & 0.20055 & 1.76150 & 0.70162 \\
$V_{7} \; \& \; V_{8} = 0$ & 0.55572 & 0.20055 & 1.76152 & 0.70244 \\
$V_{8} \; \& \; V_{9} = 0$ & 0.55572 & 0.20055 & 1.76160 & 0.70254 \\
$V_{9} \; \& \; V_{10} = 0$ & 0.55572 & 0.20055 & 1.76159 & 0.70257 \\
$V_{10} \; \& \; V_{11} = 0$ & 0.55572 & 0.20055 & 1.76162 & 0.70262 \\
$V_{11} \; \& \; V_{12} = 0$ & 0.55572 & 0.20055 & 1.76163 & 0.70261 \\
$V_{12} \; \& \; V_{13} = 0$ & 0.55572 & 0.20055 & 1.76163 & 0.70260 \\
\enddata
\tablenotetext{*}{In units of $\lambda_{0}$.}
\end{deluxetable}

Also shown in Figure 2 is the location of the stagnation point (vertical red line) and the diffusion region (salmon colored region).  The diffusion region is defined as the location where the diffusive electric field $\eta B_{a}(y)/a(y)c$ is greater than the advective electric field $Vb/c$.  The region where the reverse is true is defined as the advection region.  This region is shaded blue in Figure 2.  In the advection region the current sheet is bifurcated into slow-mode, Petschek-type shocks.

At $t = 14$, the current sheet consists almost entirely of the diffusion region.  Only near the tips of the sheet, where the current density approaches zero, does advection start to become significant.  However, this region is also where the expansion used to obtain the analytical solution breaks down.  Numerical simulations show that advection does dominate over diffusion at the tips of the current sheet, but the Petschek-type shocks are no longer present.  At the tips the outflowing plasma rapidly slows and spreads out into a larger region.

Since the field is nearly symmetric at $t = 14$, the outflow is also symmetric despite the fact that the downward directed jet encounters the solar surface, while the upward jet does not.  In general the blockage of the outflow from the lower jet causes most of the plasma flowing into the current sheet to be deflected upwards so that the downward jet is suppressed \citep{forbes86,murphy10,murphy12}.  However, if the inflow has a plasma $\beta \ll 1$, the flow becomes supermagnetosonic with respect to the fast mode-wave speed.  In this case the downward jet is not suppressed.  Instead, it is terminated by a fast-mode shock, and the flow within the current sheet remains symmetric \citep{forbes86,takasao15,zenitani15}.  The bottom panel of Figure 3 shows the fast-mode Mach number, $M_{fm}$, of the outflow as function of $y$.  The flow is supermagnetosonic in the regions where $M_{fm} > 1$.  For Petschek reconnection with an inflow plasma of zero $\beta$, the predicted value of $M_{fm}$ is $[2/(\gamma-1)]^{1/2}$ \citep{soward82, forbes86}.  For $\gamma = 5/3$ this gives $M_{fm} = 3^{1/2} \approx 1.73$, which is close to the maximum value in Figure 3.

At $t = 40$ the configuration of the fields and flows is noticeably asymmetric.  Most of the current sheet lies above the stagnation point, and there is an extended region of slow-mode shocks above the upper tip of the diffusion region.  The diffusion region itself is distributed asymmetrically around the stagnation point, although its overall length is still of order $2\alpha$, the diffusion region scale length.  Note also that the stagnation point, $y_{sp}$, lies slightly below the pinch point, $y_{pp}$ (i.e. the maximum of $B_{a}(y)$) and the neutral point, $y_{np}$, (i.e. $b = 0$) lies below the stagnation point (cf. \citealt{murphy12}).

We gain further insight into the behavior of the system by considering the analytical solutions obtained by substituting the model expression for $B_{a}$ into Equations (14) and (16).  Because the solution of Equation~(14) for the stagnation point leads to a complicated cube root, we make an additional simplifying approximation, namely that $h$ is large compared to both $\lambda$ and $y$ (i.e. $h \rightarrow \infty$).  In which case:
\begin{equation}
B_{a}(y) \approx { 2 A_{0} \lambda \sqrt{(y^2-p^2) (q^2-y^2)}\over \pi (y^2 + \lambda^2) \sqrt{(\lambda^2 + p^2) (\lambda^2 + q^2)}}.
\end{equation}
This expression provides a good approximation for $B_{a}$ in the vicinity of the lower portion of the current sheet, especially at late times.  With this approximation we obtain:
\begin{equation}
y_{sp}  \approx \sqrt{p^2\lambda^2 + q^2 \lambda^2 + 2p^2q^2 \over 2\lambda^2 + p^2 + q^2},
\end{equation}
for the location of the stagnation point and
\begin{equation}
\alpha  \approx {\sqrt{3} \; (\lambda^2 + p^2) (\lambda^2+q^2) (q^2-p^2) \over \sqrt{4(2\lambda^2 + p^2 + q^2)^3 (\lambda^2 q^2 + 2q^2p^2 + \lambda^2p^2)}},
\end{equation}
for the length scale of the diffusion region.

If the current sheet is short enough, the decrease of the external field $B_{a}(y)$ with height becomes negligible.  For such a configuration the field and flow is symmetric relative to the midpoint of the sheet, and $B_{a}(y)$ should correspond to the Green-Syrovatskii model which is of the form $B_{a}(y) = k \sqrt{L^2-y^2}$, where $k$ is a constant and $L$ is the half-length of the current sheet \citep{green65,syrovatskii71}.  When $q-p \ll p$ this condition is met, and Equation~(18) reduces to
\begin{equation}
B_{a}(y)  \approx {4 A_{0} \lambda p \sqrt{L^2-y_{*}^2} \over \pi (\lambda^2 + p^2)^2},
\end{equation}
where $L = (q-p)/2$ and $y^* = y-y_{sp}$.  For this field
\begin{equation}
y_{sp}  \approx (q+p)/2,
\end{equation}
corresponding to the midpoint of the current sheet, and
\begin{equation}
\alpha  \approx \sqrt{3} \; L/2 \approx \sqrt{3} \; (q-p)/4,
\end{equation}
which indicates that the diffusion region extends nearly the entire length of the current sheet.  The requirement that $q-p \ll p$ means that the current sheet has to be much shorter than the height, $p$, in order for the reconnection to be of the symmetric, Sweet-Parker type.  Although such a short current sheet may occur during the impulsive phase, both observations \citep{webb03,reeves11,lin15} and simulations (e.g. \citealt{linker01}) show that the current sheet in the post-impulsive phase is typically much longer than the height of the flare loops.  Therefore, we expect the falloff of the solar magnetic field with height to have a significant effect on reconnection in the post-impulsive phase.

For a long current sheet (i.e. $q \rightarrow \infty$) Equations (19) and (20) further simplify to
\begin{equation}
y_{sp}  \approx \sqrt{\lambda^2 + 2p^2},
\end{equation}
\begin{equation}
\alpha  \approx \sqrt{3/4}\; (\lambda^2 + p^2)/\sqrt{\lambda^2 + 2p^2}.
\end{equation}
We see from Equation~(24) that the altitude of the stagnation point never becomes very high.  When $p$ is small, the altitude is approximately $\lambda$, the length scale of the surface magnetic field, and when $p$ is large, it is approximately $\sqrt{2}p$, an altitude that is only slightly higher than the altitude of the flare loops.  (In fact $\sqrt{2}p$ overestimates the height.  As one can see from Table 1, the value obtained from the most accurate approximation is about 9\% smaller.)  

From Equation~(25) we also see that the diffusion region length scale, $\alpha$, is about $0.9 \lambda$ when $p$ is small and about $0.6 p$, when $p$ is large.  Thus, the length of the diffusion region predicted by this analysis is relatively large, on the order of the geometrical scale length of the field.  Despite the presence of the slow shocks, the reconnection rate, as prescribed by Equation~(5), remains close to the Sweet-Parker rate of a current sheet whose length is on the order of $\lambda$ or $p$, whichever is the larger.  Thus, the reconnection rate remains slow.  In order to have the fast reconnection we typically associate with Petschek reconnection, the diffusion region needs to be many times smaller than the global scale length of the field, but in our analysis the diffusion region remains large if the resistivity is uniform.

\section{Flare Model Dynamics}

In this section we reconsider the analytical models of \citet{reeves05} using the reconnection theory presented in the previous section.  This model prescribes a scenario for the evolution of the magnetic field shown in Figure 1(a).  This configuration develops after a loss of equilibrium is triggered by slowly pushing the source regions at $\pm \lambda$ together. At the start of the eruption, the flux rope is located close to the solar surface, and no neutral point exists below it.  As the flux rope moves upward, the neutral point appears at the surface and the vertical current sheet starts to grow.  Reconnection of field lines within this sheet causes it to detach from the surface, so that closed magnetic loops are formed below it.  The length of the current sheet is determined by how fast the flux rope moves upwards and by how fast reconnection occurs.
\begin{figure}
\epsscale{0.75}
\plotone{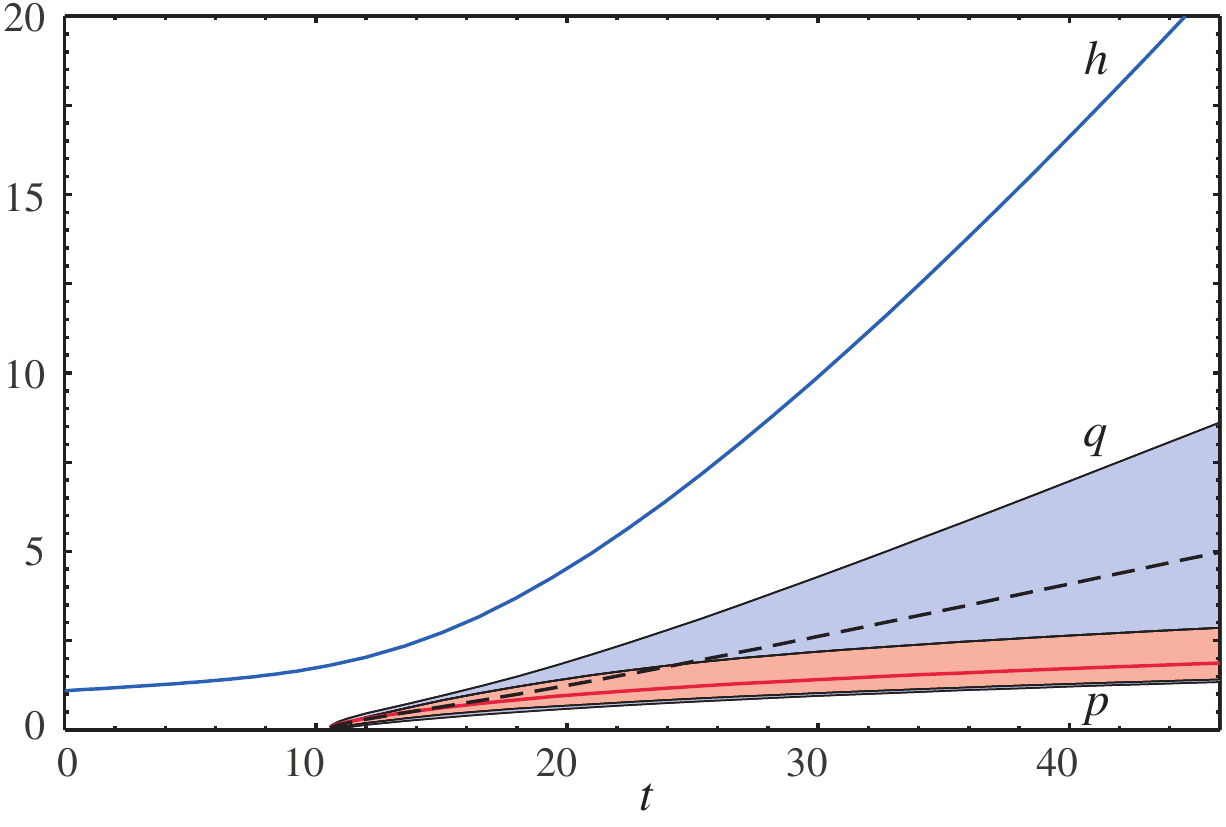}
\caption{Trajectories for the flare model of \citet{reeves05} using the more realistic reconnection model of Section 2.  The blue curve, labeled $h$, indicates the flux rope's center, while the curves labeled $q$ and $p$ indicate the paths of the upper and lower tips of the current sheet.  As in Figure 2, salmon colored shading indicates the diffusion region, and blue shading indicates the advective region, where the current sheet is bifurcated into a pair of slow-mode shocks.  The red line shows the location of the stagnation point, and the dashed line shows the stagnation-point location assumed in the original Reeves \& Forbes model.  Lengths and time are normalized to $\lambda_{0}$ and $t_{0}$, respectively (see text).}
\end{figure}

The flare model parameters $h$, $q$, $p$, and $r$, shown in Figure 1, are determined as functions of time, $t$, by invoking magnetic flux conservation, total energy conservation, and force balance within the flux rope.  Finally, to close the system of equations a prescription is needed for the reconnection rate.  The two conservation laws are based on the model's assumption that there is no injection of magnetic flux or energy during the short time scale of the eruption.  Conservation of flux yields the condition
\begin{equation}
A(0, h-r) = {A_{0}\over \pi}  \left[ 2 \ln {\left(2 \lambda_{0} \over r_{0}\right)} + {\pi \over 2} \right] ,
\end{equation}
where $A$ is the magnitude of the vector potential given by equation two in \citet{reeves05} and equation 30 in \citet{lin00}.  The right-hand side of Equation (26) is a constant.  Setting $A(0, h - r)$ to a constant means that the magnetic field is frozen at the surface of the flux rope, so field lines cannot emerge from, or be absorbed into it.  To facilitate comparison with the previous publications, the right-hand side of Equation~(26) is evaluated at the location $\lambda_{0}$ where the flux-rope current, $I$, reaches a maximum during its pre-eruption evolution.  The maximum occurs just before the flux rope reaches the critical point, so that $\lambda$ at the time of eruption is slightly less than $\lambda_{0}$.  The constant $r_{0}$ is the radius of the flux rope at $\lambda = \lambda_{0}$.

Conservation of energy requires
\begin{equation}
W_{\rm{ME}} + W_{\rm{KE}} + W_{\rm{TE}} =  {\left(A_{0} \over \pi \right)}^2 \; \left[ \ln {\left(2 \lambda_{0} \over r_{0}\right)} + {3 \over 4} \right],
\end{equation}
where $W$ indicates energy per unit length.  The right-hand side of Equation~(27) is the magnetic energy of the configuration per unit length at $\lambda = \lambda_{0}$ (see equation 15 in \citealt{reeves05}).  $W_{\rm{KE}}$ is the kinetic energy of the flux rope per unit length, namely
\begin{equation}
W_{\rm{KE}} =  m_{f} \; \dot{h}^2 / 2,
\end{equation}
where $m_{f}$ is the mass of the flux rope per unit length and $\dot{h}$ is the flux rope velocity.  $W_{\rm{TE}}$ is a measure of the energy per unit length that is available for heating the flare plasma.  It is defined as the Poynting flux into the current sheet integrated over time and the length of the sheet.  That is
\begin{equation}
W_{\rm{TE}} =  {c \over 2\pi} \int^q_{p}{\int^t_{t_{n}} {E B_{a} \; dt \; dy}}.
\end{equation}
\begin{figure}
\epsscale{0.7}
\plotone{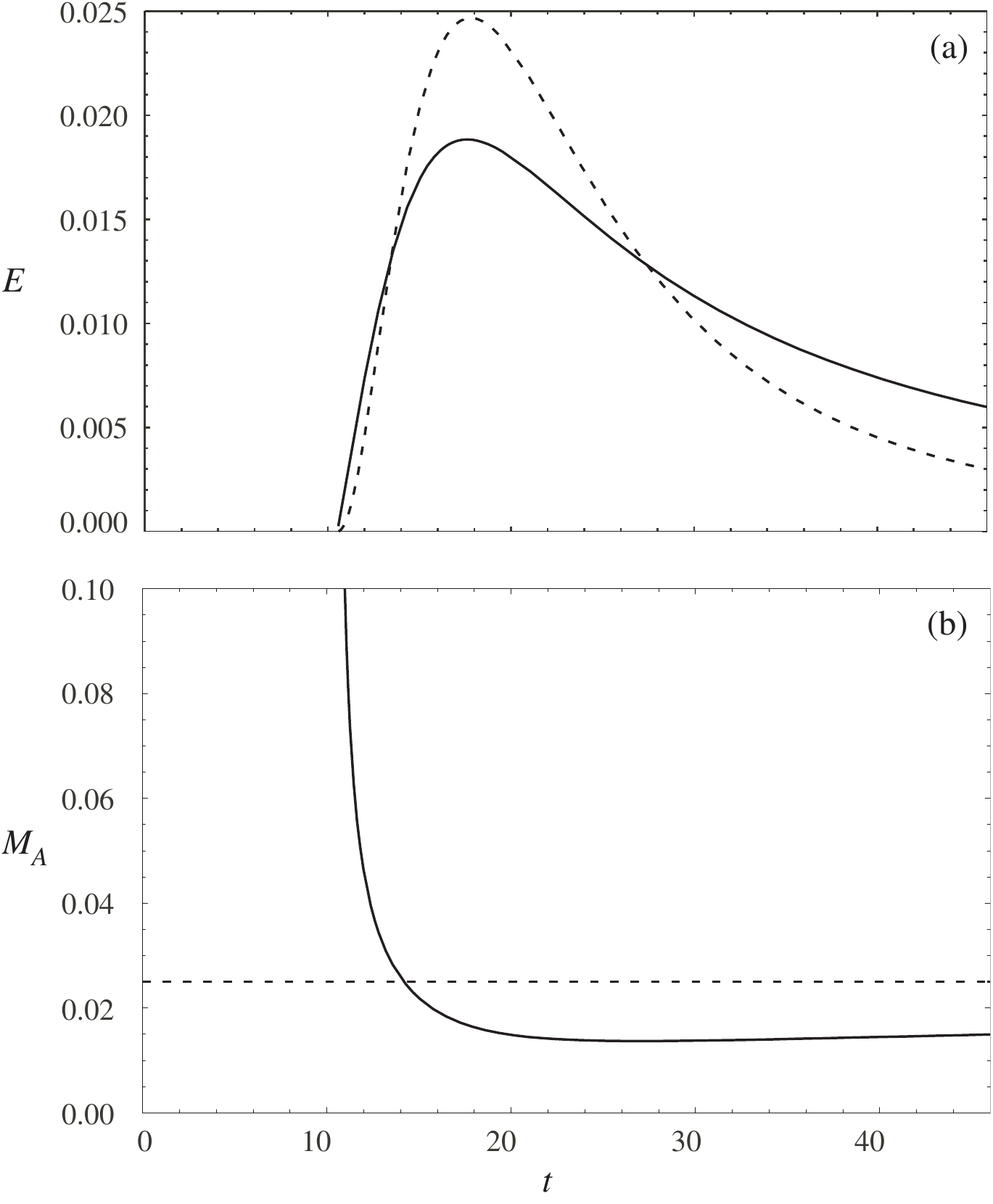}
\caption{The reconnection electric field, $E$, (panel a) and inflow Alfv\'{e}n Mach number, $M_{A}$,  (panel b) at the stagnation point as functions of time.   The dashed curves show the values for the original Reeves \& Forbes model.  The electric field is normalized to $A_{0}V_{0}/(\lambda_{0}c)$.}
\end{figure}

The free magnetic energy, $W_{\rm{ME}}$, is computed by calculating the work done by the flux rope during the eruption.  The force per unit length on the flux rope can be expressed as the sum of an internal force and an external force
\begin{equation}
{\bf F_{B}} (h) = {1 \over c} \int \! \! \int {\bf j_{f}} \times {\bf B_{f}} \; d \sigma + {1 \over c} \int \! \! \int {\bf j_{f}} \times {\bf B_{e}} \; d \sigma.
\end{equation}
Here $\sigma$ is the region occupied by the flux rope, ${\bf j_{f}}$ is the flux-rope current density, and ${\bf B_{f}}$ and ${\bf B_{e}}$ are the magnetic fields due to the internal current of the flux rope and the external currents outside the flux rope, respectively.  We assume $\sigma$ is small enough to make the external field ${\bf B_{e}}$ effectively uniform within the flux rope.  We also assume that if $\sigma$ is sufficiently small, the internal configuration of the flux rope will remain close to an equilibrium state during the eruption.  With these assumptions the internal state of the flux rope satisfies the force-free field condition
\begin{equation}
{\bf j_{f}} \times {\bf B_{f}} =  {\bf 0} ,
\end{equation}
and the external force that accelerates the flux rope upwards is prescribed by
\begin{equation}
 {\bf F_{B}} =  IB_{e}{\bf \; \hat{y}}/c .
 \end{equation}
 \begin{figure}
\epsscale{0.7}
\plotone{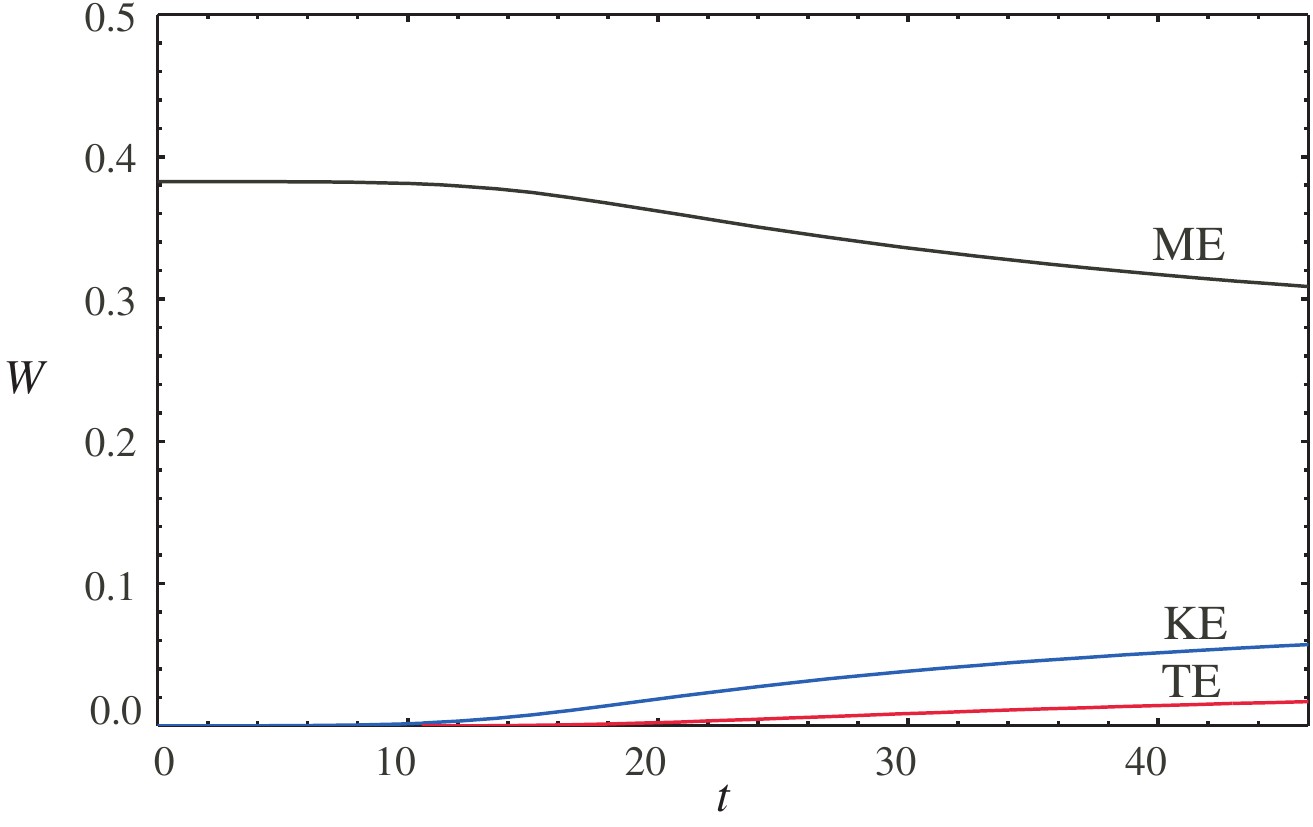}
\caption{Free magnetic energy (black curve labeled ME), flux-rope kinetic energy (blue curve labeled KE), and thermal flare energy (red curve labeled TE) per unit length as functions of time.}
\end{figure}
The free magnetic energy is, then
\begin{equation}
W_{\rm{ME}} =  {1 \over c} \int^\infty_{h} {I(h') B_{e}(h') \; dh'},
\end{equation}
where $I$ is given by Equation~(2).  Integration of Equation~(31) over the area of the flux rope leads to the condition \citep{isenberg93,forbes95}
\begin{equation}
r  I \approx  r_{0}I_{0},
\end{equation}
where $I_{0}$ is the current at $\lambda = \lambda_{0}$, when the current is at its maximum value.
\begin{figure}
\epsscale{0.7}
\plotone{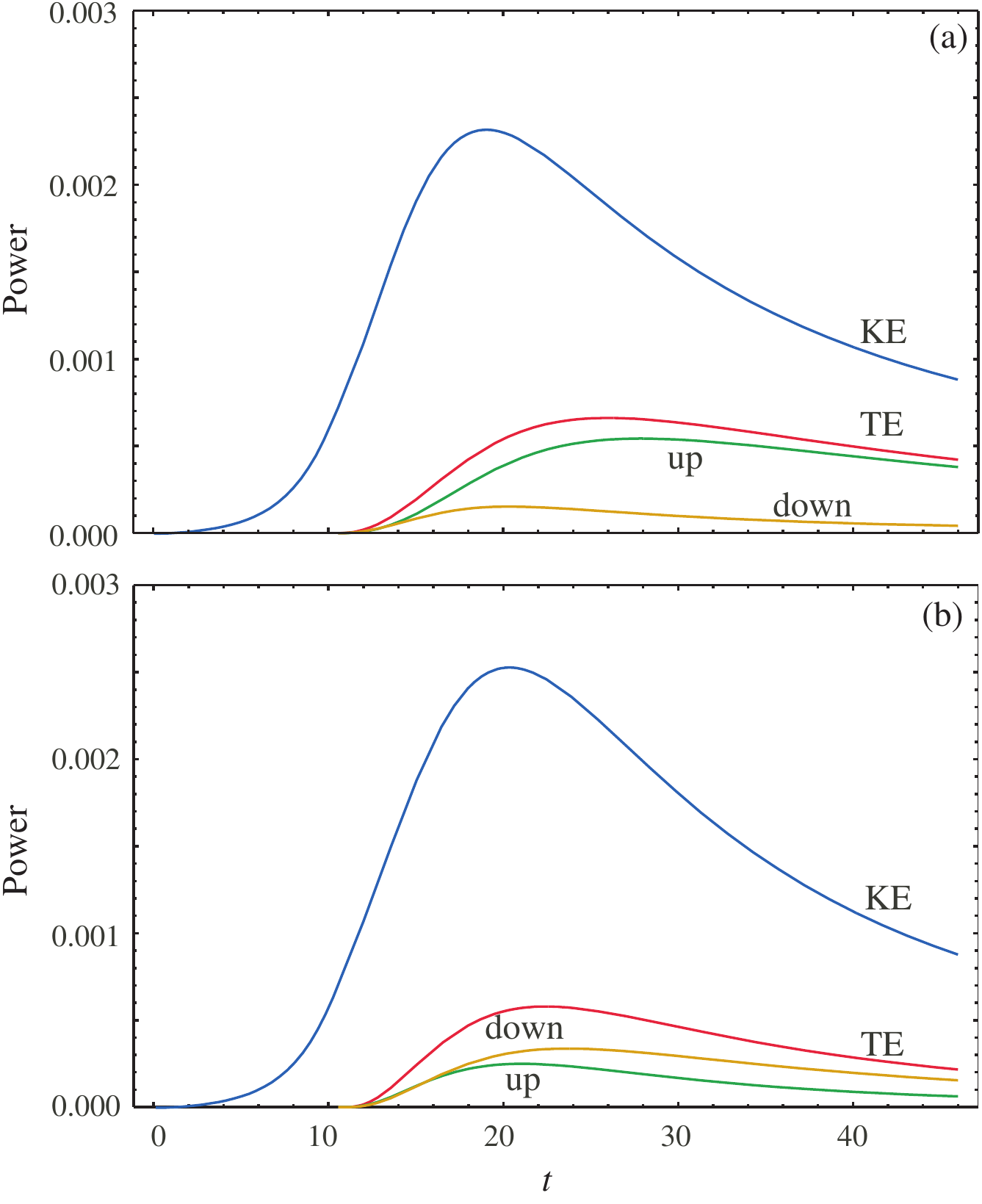}
\caption{Power output per unit length versus time for both the revised model (panel a) and the original Reeves \& Forbes model (panel b).  The blue curves show the kinetic power, and the red curves show the thermal flare power output.  The green and yellow curves show the portions of the thermal flare power in the upper and downward directions.  Because of the lower altitude of the stagnation point in the revised model, the amount of thermal power in the downward flow is noticeably lower than in the original model.  Power/length is in units of $A_{0}^2 / t_{0}$.}
\end{figure}

Finally, to close the system of equations we need to prescribe the electric field in the current sheet as a function of time.  To lowest order in the expansion, this electric field is uniform within the current sheet.  Using Faraday's equation we can write it in terms of the magnitude of the vector potential as
\begin{equation}
\partial A_{cs} / \partial t = -\; cE_{cs},
\end{equation}
where $A_{cs}$ and $E_{cs}$ are the values of $A$ and $E$ at $x = 0$ for $p < y < q$.  The model of \citet{reeves05} arbitrarily assumes that
\begin{equation}
cE_{cs} = M_{A0} B_{a}^2(y_{1/2})(4\pi \rho_{a})^{-1/2},
\end{equation}
where $y_{1/2} = (p+q)/2$ is the midpoint of the current sheet, $M_{A0}$ is the inflow Alfv\'{e}n Mach number at $y_{1/2}$, and $ \rho_{a}$ is the ambient plasma density of the corona.  In the Reeves \& Forbes model, $M_{A0}$ is a free parameter that is constant in time and in the range between 0 and 1.  We now replace this ad-hoc expression with one that is based on the physical reconnection model of the previous section.  Replacing $M_{A0}$ by $M_{A}$ from Equation~(5) and $y_{1/2}$ by $y_{sp}$, we obtain
\begin{equation}
cE_{cs} = (\eta/\alpha)^{1/2} B_{a}^{3/2}(y_{sp}) (4\pi \rho_{a})^{-1/4},
\end{equation}
where to the lowest order of approximation, $y_{sp}$ and $\alpha$ are given by Equations~(19) and (20).  Because the lowest order approximation leads to significant errors ($\sim 30$ \%) when the configuration is highly asymmetric, we use interpolating functions obtained by setting 
$V_{3}$ and $V_{4}$ to zero in place of Equations~(19) and (20).  The improved accuracy in the calculation of $\alpha$ and $y_{sp}$ is shown in Table 2.
\begin{figure}
\epsscale{0.7}
\plotone{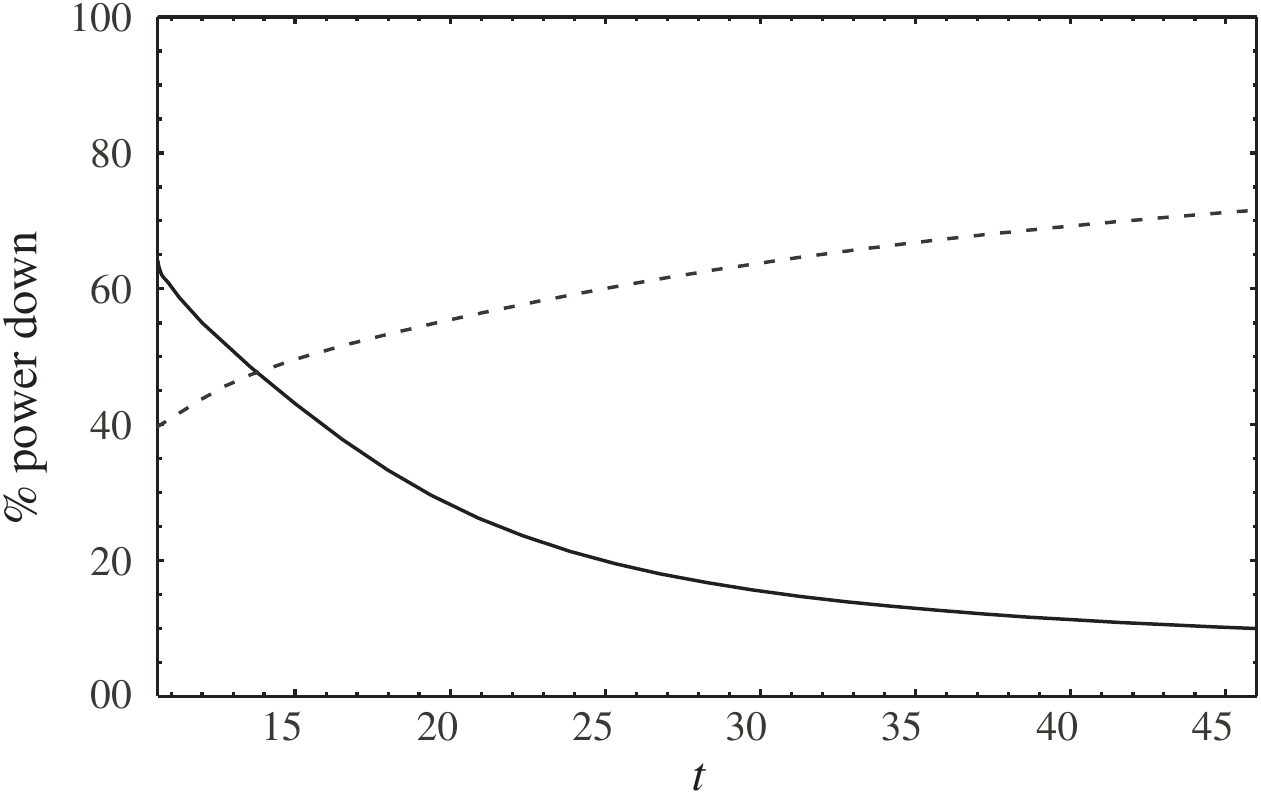}
\caption{Percentage of total thermal power directed downward as a function of time for the revised flare model (solid curve) and the original flare model (dashed curve).  The lower percentage in the revised model at late times is caused by the much lower position of the stagnation point.}
\end{figure}

Equations~(2), (26), (27), (34), and (35), together with the subsidiary Equations~(28), (29), (33), and (37), determine the evolution of the flare model parameters $I$, $h$, $q$, $p$, and $r$ as functions of time.  To obtain a specific solution, we need to specify the magnetic diffusivity, $\eta$.  This quantity can be expressed in terms of the dimensionless Lundquist number
\begin{equation}
\mathrm{Lu} = V_{A\lambda} {\lambda_{c} \over  \eta} = {A_{0} \over \pi \eta} \; {1 \over \sqrt{4 \pi \rho_{a}}},
\end{equation}
where  $V_{A\lambda} = B_{\lambda}/(4 \pi \rho_{a})$, $B_{\lambda} = A_{0} / (\pi \lambda_{c})$, $\lambda_{c}$ is $\lambda$ at the loss of equilibrium point, and $\rho_{a}$ is the ambient plasma density in the corona.  With this definition Lu is invariant during the eruption, whereas the more standard definition based on the length of the current sheet is not.

\begin{deluxetable}{ccccccc}
\tablecaption{Comparison of Errors Produced by Interpolated and Analytical Approximations (Values are Normalized to $\lambda_{0}$)}
\tablecolumns{7}
\tablewidth{0pt}
\tablehead{
\colhead{} &
\colhead{$t$} &
\colhead{Precise\tablenotemark{a}} &
\colhead{Interpolated\tablenotemark{b}} &
\colhead{Interp. Error} &
\colhead{Analytical \tablenotemark{c}} &
\colhead{Anal. Error}}
\decimals
\startdata
$\alpha$ & 14 & 0.2006 & 0.2001 & \phs 0.3 \% & 0.2076 & \phs 3.5 \% \\
$y_{sp}$ & 14 & 0.5557 & 0.5284 &$-4.9$ \% & 0.5410 & $-2.6$ \% \\
$\alpha$ & 40 & 0.7026 & 0.6826 & $-2.8$ \% & 0.9443 & \phd 34.4 \% \\
$y_{sp}$ & 40 & 1.7596 & 1.7171 & $-2.4$ \% &  1.9131 & \phs 8.7 \% \\
\enddata
\tablenotetext{a}{ from setting $V_{12} \; \& \; V_{13}$ to zero}
\tablenotetext{b}{ values from fitting surface in $p$-$q$ space to the $V_{3} \; \& \; V_{4}$ solutions with $h \rightarrow \infty$}
\tablenotetext{c}{ from setting $V_{2} \; \& \; V_{3}$ to zero with $h \rightarrow \infty $}
\end{deluxetable}

Figure 4 shows the trajectories obtained for $r_{0} / \lambda_{0} = 0.1$, $\rho_{a} / \rho_{0} = 6.46 × 10^{-5}$, $m_{0}/(\lambda_{0} m_{f}) = 4$, and Lu = 18517 where $m_{0}$ is the mass of the flux rope.  The most noticeable difference between these trajectories and the previous ones of \citet{reeves05} is the low altitude of the stagnation point (red line).  At late times it lies just above the top of the flare loops rather than at the midpoint $(q + p)/2$ (dashed line).  The location of the neutral point is even lower, since it lies below the stagnation point (cf. Figure 2).  The lengths in Table 2 and Figure 4 are normalized to $\lambda_{0}$.  The time in Figure 4 is normalized to a scale time based on the length, $\lambda_{0}$, and the velocity $V_{0} = B_{0} \rho_{0}^{-1/2} = A_{0} (\lambda_{0} / m_{0})^{1/2}$.

Figure 5 shows the corresponding reconnection rate as measured by the electric field (Figure 5a) and the inflow Alfv\'{e}n Mach number (Figure 5b) at $y_{sp}$.  The dashed curves show the results obtained by \citet{reeves05} using a constant inflow Alfv\'{e}n Mach number of 0.025.  The solid curves show the results of the new reconnection model.  The new model also contains a free, or loosely specified, parameter, namely Lu, so we need to be careful when comparing these two models to distinguish between the physical differences of the models and those caused by using different reconnection rates.  In order to do this we select a value of Lu = 18517, so that the amount of reconnected flux at the last time shown in Figure 4 (i.e. $t = 46$) is the same for both the old and new models.  The effect of this constraint is to force the area under the curves for $E$ to be the same.  The principal difference between the new model and the old one is that the inflow Alfv\'{e}n Mach now varies with time.  $M_{A}$ is very large when the magnetic neutral point first appears at $t = 10.6$, and then drops rapidly to a nearly constant value of about 0.0315 by $t = 20$.  Thus, the assumption of the previous model that $M_{A}$ is roughly constant is a reasonably good approximation during the late phase of the evolution.  The main deficiency of the old model is that it places the reconnection site at too high an altitude.  This higher position also causes the reconnection site to propagate upwards at too fast a speed.   In the new model the reconnection site is always located a relatively short distance above the top of the flare loops, and it propagates upward at roughly the same speed that they do.

One of the main goals of the previous work by Reeves \& Forbes was to determine the energy output predicted by the two-dimensional model as a function of time.  Figures 6 and 7 show the energy and power output by the new model for the same parameters used in Figures 4 and 5.  The decrease in the free magnetic energy (ME) shown in Figure 6 is essentially the same as before, but the "thermal" flare energy release (TE) is about double the old one.  (Recall that TE is the integrated Poynting flux into the current sheet.)  This increase is due to the fact that the reconnection site in the new model remains at low altitude rather than rocketing up to high altitudes as before.  The magnetic field at the lower altitude is significantly stronger than at the higher one, so the Poynting flux is now greater than before.  However, the percentage of this Poynting flux that is channeled downward is much smaller than before because most of the current sheet now lies above the neutral and stagnation points.  Even though the thermal energy has doubled, the amount of this energy channeled downward is so reduced that the net downward energy is less than half of what it was before.
\begin{figure}
\epsscale{0.87}
\plotone{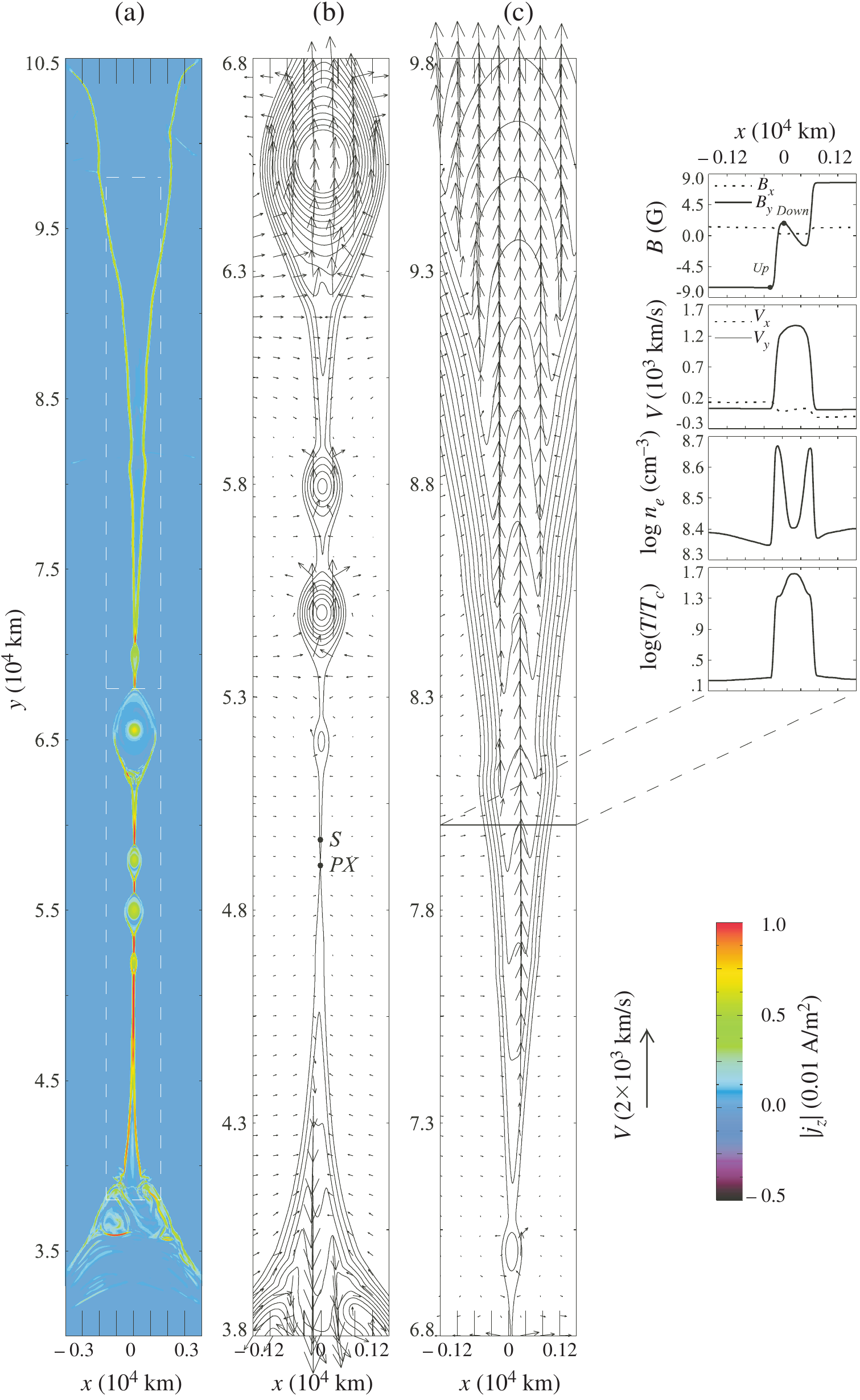}
\caption{Plots of the magnetic field lines, flow vectors and current density (color scale) in an MHD simulation of the eruptive flare model with a large numerical domain.  At the time shown, the current layer in the upper portion of the box has bifurcated into slow-mode shocks despite the formation of magnetic islands lower down (after \citealt{mei12}).}
\end{figure}
The effect of the new model's low altitude neutral point is shown in Figure 8.  At early times the percentage of power directed downward in the new model is about 60\% compared to 40\% for the old model, but these percentages rapidly reverse.  By $t = 46$ only about 15\% of the thermal power is directed downward.  This low value significantly reduces the energy channeled into the flare ribbons, a reduction that favors the estimate by \citet{klimchuk96} that only about one percent of the energy released by magnetic reconnection is needed to create the flare ribbons.  However, it should be kept in mind that the distinction between upward and downward directed energy flows becomes somewhat moot in three-dimensions.  In the fully three-dimensional versions of this model \citep{titov99,isenberg07,kliem12} all of the field lines remain attached to the solar surface so that distinction between up and down transfers into a distinction between the energy transferred to different regions of the solar surface.

A key feature of the new reconnection model is the prediction of slow-mode shocks lying above the stagnation point of the flow during the post-impulsive phase.  A simulation by \citet{mei12} that does in fact exhibit such shocks is shown in Figure 9.  At late times, when the current sheet has become quite long, an extended set of slow-mode, Petschek-type shocks are seen above the stagnation point.  Other simulations do not typically see these shocks because their current sheets are too short.  Unfortunately, quantitative comparison between the new reconnection model and the Mei et al. simulation is not possible for two reasons.  First, the Mei et al simulation uses a density model that decreases with height, whereas the model presented in Section 2 does not.  Second, the simulation uses a Lundquist number on the order of $10^4$.  At this value the assumption of laminar flow starts to break down.  For values of Lu greater than about $10^4$ the Sweet-Parker diffusion region becomes unstable to tearing \citep{loureiro07,tenerani16}, and by the time shown in Figure 9, numerous magnetic islands have started to form.  Their appearance causes the external field, $B_{a}$, to deviate markedly from the simple form given by Equation~(17).  The formation of islands inhibits the formation of extended, slow shocks \citep{innocenti15}.  Nevertheless, the tendency for the upper set of slow shocks to form is still evident.

 Another feature of the reconnection model of Section 2 that is supported by the \citet{mei12} simulation is the location of the stagnation point (indicated by "$S$" in panel (b) of Figure 9).  Despite the presence of multiple neutral points, only a single stagnation point occurs in the current sheet.  The presence of a single stagnation point means there is just one upward directed jet and one downward directed jet.  These outflows are produced by a principal neutral point (indicated by "$PX$" in panel (b) of Figure 9 that dominates the dynamics of the current sheet.  The location of the stagnation point, and of the principal neutral point associated with it, are within about 20\% of the location predicted by the reconnection model.  At the time shown in Figure 9, the top of the flare loop system is at $y = 3.7$.  Based on Equation~(24) we would expect $y_{sp}$ to occur at a height of about 6 in Figure 9.  This value is somewhat larger than the 5.0 that actually occurs in the simulation.  In any case it is clear that at late times the stagnation point and associated neutral point do not occur at the midpoint of the current sheet as assumed by \citet{reeves05}.

\section{Relevance to Observations}

Within the last ten years observations of current sheets formed in the wake of erupting solar flares have greatly improved \citep{ciaravella08, lin15, reva16, seaton17}.  Within the current sheets small features are sometimes observed that move at high speeds (100 to 500 km s$^{-1}$) in a manner suggestive of reconnection outflow jets \citep{savage10,kumar13,takasao12}.  The true nature of the features remains unknown at the present time.  Some features appear to be regions of low density with a three-dimensional, loop-like geometry \citep{savage12}, while other features appear to be regions of enhanced density that look more like magnetic islands.  The low density, downward moving features also generate oscillatory wakes that may be due to a Raleigh-Taylor-type instability \citep{guo14, innes14}.

Particularly intriguing are the moving features observed by the X-Ray Telescope (XRT) on {\it Hinode} for an eruption that occurred on 2008 April 9.  This event, known as the "Cartwheel" event, produced an extended current-sheet like structure that lasted for many hours \citep{savage10}.  Within this structure small features could be seen moving downward at low altitudes and upwards at high altitudes as shown in Figure 10(a).  The movement of these features was quite rapid, ranging between 80 to 180 km s$^{-1}$, a speed that is much faster than the slow, upward motion ($< 2$ km s$^{-1}$) of the flare loop system.  A remarkable aspect of the features is that they are already moving at their maximum velocity the moment they are first observed.  The only obvious change in speed occurs in the downward moving features, which decelerate as they approach the top of the flare loop system.  The minor fluctuations that are seen in the position of the features with time are most likely due to observational uncertainties.  The upward moving features do not show any change in speed in so far as one can tell from the few observations that are available.  It is possible to follow some individual upward moving features from the XRT field of view into the field of view of the Large Aperture Solar Coronagraph (LASCO) on the Solar Heliospheric Observatory \citep{savage10, schanche16}.

If we assume that the features move with the plasma, then their motion implies the existence of downward and upward directed jets with nearly constant velocity within the current sheet.  Furthermore, the region where the jets are accelerated must be shorter than the resolution limit of the XRT.  For such faint, rapidly moving features this limit is on the order of $10^4$ km.  Evidence for a short, sub-resolution diffusion region is also implied by the observation of the trajectories in Figure 10(a) labeled "disconnection event".  Here two density-enhancement features simultaneously appear very close to one another, but one moves upwards, while the other moves downwards.  Because of three-dimensional projection effects, it is difficult to obtain an accurate estimate of the distance between the two features when they first appear, but it is probably less than $10^4$ km.

The observed flow within the current sheet more closely resembles what we expect to see for Petschek reconnection rather than Sweet-Parker reconnection.  If the entire sheet were a simple Sweet-Parker current sheet we would expect to see flows steadily accelerating from zero at the stagnation point to something close to the ambient Alfv\'{e}n speed at the tips of the current sheet.  Furthermore, we would expect the stagnation point to occur in the middle of the current sheet and to propagate rapidly upwards as the current sheet lengthens in time.  Instead we see what suggests a very small diffusion region located near the lower tip of the current sheet and just above the flare loops.  The apparent upward motion of the inferred diffusion region is similar to that of the flare loops themselves (cf. Figure 4).
\begin{figure}
\epsscale{0.65}
\plotone{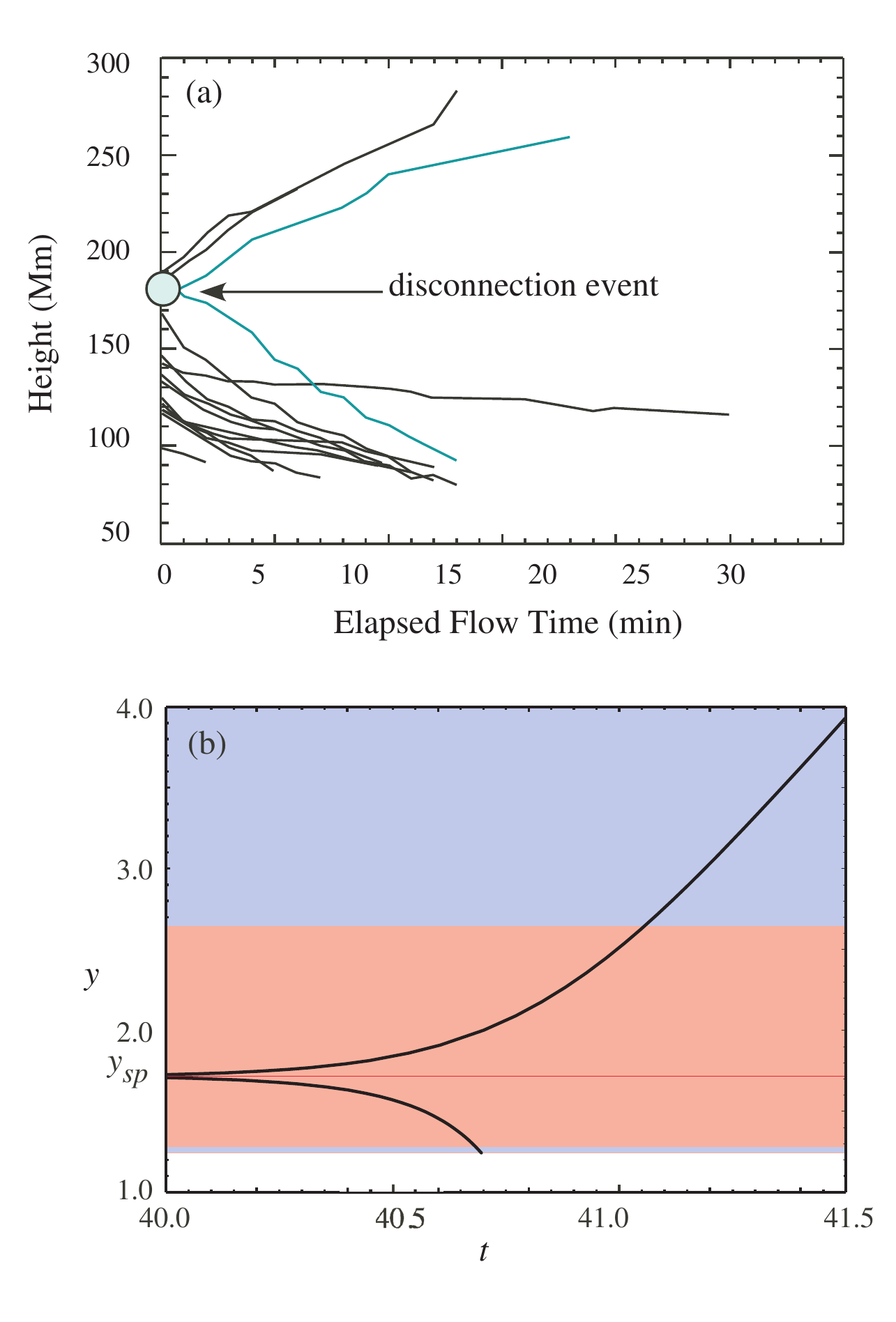}
\caption{(a) Downward and upward outflow features observed by the X-Ray Telescope (XRT) on {\it Hinode} for an eruptive flare on 2008 April 9 (after \citealt{savage10}). (b) Characteristic paths of the fluid elements for the reconnection outflows shown in Figure 2.  In the diffusion region (salmon shading) the fluid elements accelerate smoothly from zero up to a speed close to that of the ambient Alfv\'{e}n speed over an extended region.  By contrast the flow features seen in the observations show no indication of any acceleration in the region where they are observed.  This behavior suggests that the actual diffusion region is much shorter than that predicted by a model with uniform resistivity.}
\end{figure}

Despite some similarities, the trajectories in Figure 10(a) do not really match the expected trajectories from the reconnection model discussed in the previous sections.  Although the model predicts the existence of Petschek-like shocks above the stagnation point, it predicts a rather lengthy diffusion region.  Below the stagnation point this diffusion region extends all the way to the lower tip of the current sheet.  Thus below the stagnation point, the model predicts that we should see the flow being accelerated as it moves from the stagnation point to the lower tip of the current sheet as shown in Figure 10(b).  An even lengthier acceleration region is predicted to occur above the stagnation point.  Since there is no indication of such regions in the observations, we conclude that it is unlikely that the length of the diffusion region is determined solely by the geometry of the magnetic field as the model assumes.  Some additional physical process is needed to create a much smaller diffusion region.  Two likely candidates are the existence of a nonuniform resistivity and the onset of turbulence within the current sheet.  First we consider the possibility of nonuniform resistivity.

Many analytical and numerical treatments of magnetic reconnection assume that the resistivity is uniform and constant in time.  There is, however, no physical theory to support this assumption.  The assumption of uniformity is usually made for reasons of simplicity and because there is no generally accepted method for calculating the flare plasma's resistivity.  Reliable resistivity formulas do exist for collisional plasmas \citep{spitzer62, braginskii65}, but these are unlikely to be valid in the low density, high-electric-field environment of a flare \citep{holman85}.

Several simulations have been done using hypothetical, anomalous resistivity models.  \citet{ugai07} and \citet{yokoyama01} have used anomalous resistivity models of the form $\eta = k_{d} (V_{d} - V_{c})$ for $V_{d} > V_{c}$, and $\eta = 0$ for $V_{d} \leq V_{c}$.  Here $V_{d} = |j / \rho|$, is the electron drift speed, $j$ is the current density, $k_{d}$ is a constant, and $V_{c}$ is a threshold velocity for the onset of a current-driven instability.  Another model that has been used is $\eta = k_{j} (j-j_{c})^2$ for $|j| > j_{c}$, and $\eta = 0$ for $j \leq j_{c}$.  The parameter $k_{j}$ is a constant, and again $j_{c}$ is a threshold for the onset of the instability that creates the anomalous resistivity.  Since the parameters $k_{d}$, $k_{j}$, $V_{c}$, and $j_{c}$ are poorly known, these models do little to constrain the values of $\eta$ that might occur in flares.  However, they do provide a way to localize the resistivity to a small region.  The value of $j$ within the flare current sheet has its maximum value close to the pinch-point, so by setting the value of $V_{c}$ or $j_{c}$ to an appropriate value, one can confine the anomalous resistivity to a small region.  If we were to incorporate such a mechanism into the analysis of Section 2, then the principal effect would be to shorten the length of the diffusion region without significantly changing the location of the stagnation point (cf. \citealt{baty14}).  (Note that for a localized resistivity the nozzle equation we use here breaks down if the diffusion region is too short.  See Appendix A in \citealt{forbes13}.)

Two-fluid MHD theory provides a different approach to enhancing resistivity in a localized region.  This theory includes the additional effect of the Hall term, ${\bf j \times B}/nec$, where $n$ is the particle density and $e$ is the electron's charge.  The presence of this term can lead to rapid reconnection with an effective diffusion region whose size is on the order of the ion-inertial length \citep{ma96}.  As in the Petschek model, the current sheet outside the diffusion region is bifurcated, although here the bifurcation is due to whistler waves rather than slow-mode waves \citep{cassak05}.

The importance of the Hall term and other kinetic effects for flare reconnection is difficult to assess.  A justification for including it is the fact that a flare's inductive electric field is many orders of magnitude greater than the Dreicer electric field \citep{holman85, qiu02}.  The existence of such a strong electric field means that particle collisions within the flare plasma are not frequent enough to prevent the generation of runaway electrons.  Consequently, kinetic effects such as the Hall term become important.  On the other hand, the ion-inertial length in the corona is only about 10~meters, more than six orders of magnitude smaller than the scale size of a large flare ($> 10^4$ km).  The small-scale structure of the Hall diffusion region with its associated whistler waves is not stable over such a large scale \citep{daughton06}.  So it seems unlikely that the large-scale current structures of flares are directly produced by micro-scale kinetic processes.

Large-scale, MHD turbulence is another mechanism that can localize the diffusion region.  Analytical studies and numerical simulations have established that the simple Sweet-Parker current sheet is unstable to magnetic tearing when the Lundquist number, $S$, exceeds $ \sim 10^4$ \citep{loureiro07,bhattacharjee09,tenerani16}.  Since the inflow Alfv\'{e}n Mach number, $M_{A} = S^{-1/2}$ in Sweet-Parker theory, the Sweet-Parker current sheet is unstable for any value of $M_{A}$ less than about 0.01.  Once instability occurs, the current sheet no longer consists of a single sheet whose narrow width restricts the plasma flow.  Instead, it consists of large-scale magnetic islands that permit a much greater flow of plasma through the sheet.  Consequently, the length of the diffusion region at the principal neutral point (cf. Figure 9) is limited to a relatively short region within the current sheet, much as it is in Petschek reconnection.  A simulation by \citet{shibayama15} shows localization of the diffusion region by a combination of magnetic islands and Petschek-type shocks.

An attractive feature of the turbulence model is that it also provides a possible explanation for why moving features are seen within the current sheet \citep{mckenzie13}.  Although the exact nature of these features is not fully understood \citep{schanche16}, it is tempting to think they are the three-dimensional equivalents of the magnetic islands that occur in the \citet{mei12} simulation (see also \citealt{barta08}).  Figure 11 shows the trajectories of these islands within the current sheet.  The simulation trajectories close to the stagnation point at $y_{sp}$, have the expected shape for a flow that is being accelerated.  As the islands move out of the diffusion region, their trajectories become more linear, which is also as expected.  However, many of the islands do not form until the flow in which they are embedded, is outside the diffusion region.  Recently, \citet{mei17} have completed a fully three-dimensional simulation of the eruptive flare model.  In this 3D simulation an extended set of slow shocks still forms above the diffusion region, but the islands in the lower part of the current sheet are replaced by flux tubes that extend out of the plane of Figure 9.  The tubes form distorted cylinders that meander within the plane of the current sheet (see figure 4 in \citealt{mei17}).

Another mechanism that might be responsible for creating a short diffusion region is viscosity.  A simulation by \citet{baty09} demonstrates that a nonuniform viscosity can create a Petschek-type configuration even when the resistivity is uniform.  It may be possible to incorporate such a viscosity into Equation (3), but the details of how to do this have yet to be worked out.  The transport of momentum by viscosity into the upstream region can create a double-layered structure because the thickness of the current layer and the outflow layer need not be the same \citep{craig05, reeves10, craig12}.

\section{Conclusions}

The analytical flare model considered in this paper contains a feedback loop between a loss-of-equilibrium mechanism and magnetic reconnection.  Slow evolution of magnetic sources at the solar surface causes a coronal flux rope to lose its equilibrium.  Once equilibrium is lost, the flux rope is ejected upwards, and a vertical current sheet forms beneath it.  Reconnection acts to remove the current sheet and to liberate the free magnetic energy associated with the flux rope's current.  Without reconnection the flux rope cannot escape and the amount of energy liberated is on the order of 1\%, or less \citep{forbes94}.  Without the loss of equilibrium a current sheet never forms, and reconnection never occurs.
   
Previous incarnations of the model (e.g. \citealt{lin00,reeves05}) treated the reconnection in an ad hoc manner by simply assuming that the inflow Alfv\'{e}n Mach number at the midpoint of the current sheet was constant in time.  The constant was treated as a free parameter that could be adjusted to match observations.  Here we have replaced this ad hoc treatment with one that is based on physical principles.  These principles are distilled into the reconnection-nozzle Equation~(3).  This equation was first derived in its incompressible form ($\beta \rightarrow \infty$) by \citet{vasyliunas75} and extended to include compressible plasmas by \citet{titov85a,titov85b}.  Although the equation has been known for sometime, only within the last few years has it been understood how to apply it to actual problems \citep{forbes13,baty14}.
\begin{figure}
\epsscale{0.7}
\plotone{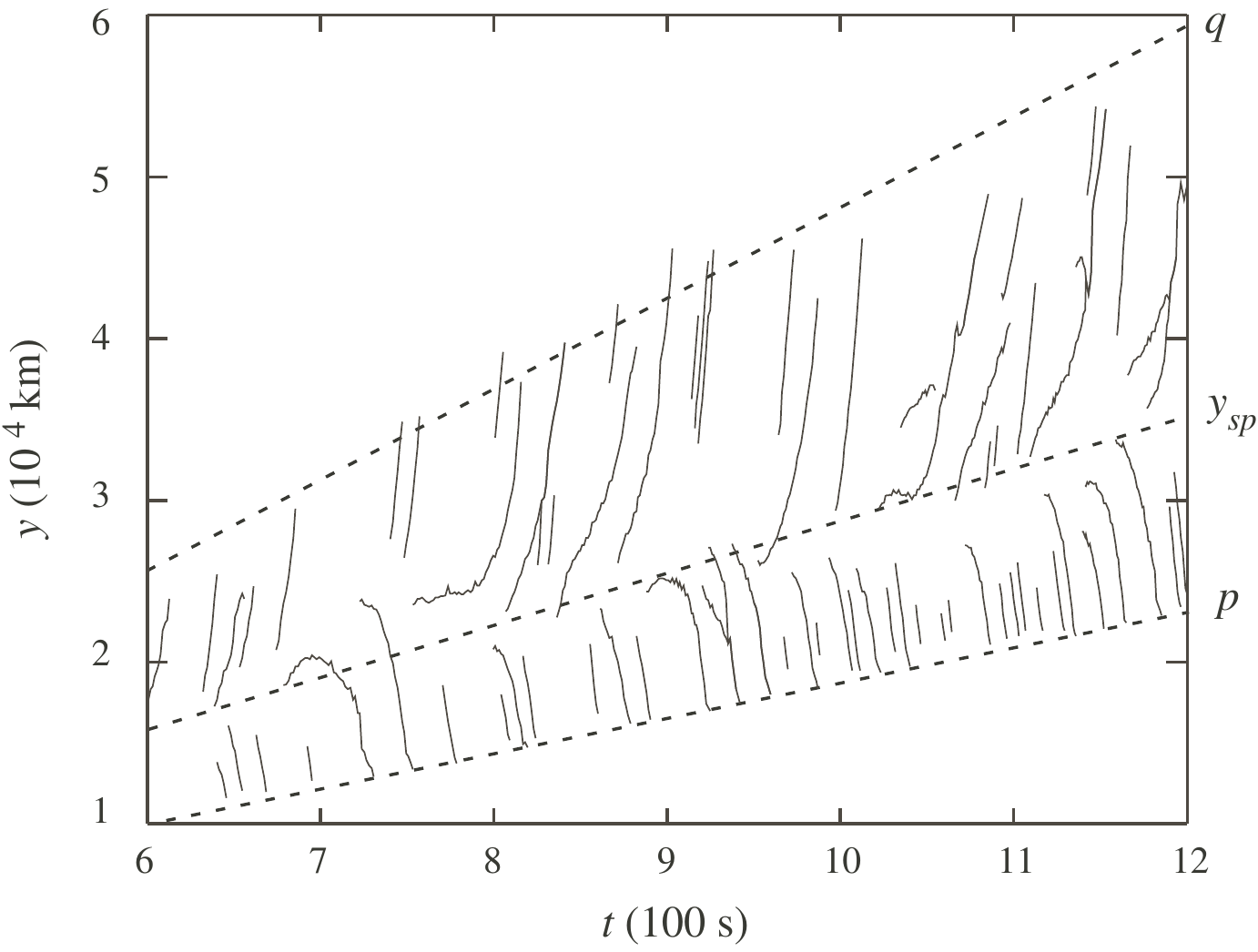}
\caption{Trajectories of individual magnetic islands (solid lines) within the current layer of the simulation shown in Figure 9 (\citealt {mei12}, Fig 13a).  The three dashed lines show the average location of the upper $y$-point ($q$), the lower $y$-point ($p$), and the stagnation point ($y_{sp}$).}
\end{figure}

A significant difference between the new reconnection model and the old one is the location of the neutral point.  Now it is located near the lower tip of the current sheet, just above the flare loops, instead of at the sheet's midpoint.  The neutral point and nearby stagnation point are located slightly below the pinch-point of the magnetic field (cf. Equation~24).  Another important difference is that Petschek-type, slow-mode shocks appear above the neutral point during the post-impulsive phase of the eruption.  However, under the assumption of uniform resistivity and laminar flow, the reconnection remains slow (cf. Equations 5 \& 25).  This slowness is due to the fact that Sweet-Parker diffusion region remains large, on the order of the height of the flare loops.  Thus, despite the presence of slow-mode shocks, the reconnection rate is closer to the slow Sweet-Parker rate than the fast Petschek rate.  Our results imply that the key to obtaining fast reconnection lies in reducing the length of the diffusion region.  One way the reduction might be accomplished is for the resistivity to be enhanced in the region where the current density is its strongest.  Another way is for the diffusion region to become unstable and turbulent when its length exceeds a critical length.

It is possible, at least in principle, to distinguish the diffusion region from the rest of the current sheet by measuring the velocity of the outflowing plasma as function of distance.  Within the diffusion region plasma accelerates from zero at the stagnation point up to a maximum speed on the order of the Alfv\'{e}n speed of the ambient corona.  In the rest of the current sheet the velocity is constant or decreases slightly as the tips are approached (Figure 2).  Observations of flow features within a current sheet observed by the XRT on {\it Hinode} show no indication of an acceleration region.  Flow features are already moving at a nearly their maximum velocity as soon as they are detected.   The only changes in speed that are observed are the deceleration of the downward directed flows as they approach the top of the flare loops (Figure 10a).  We infer, therefore, that the diffusion region must be smaller than the resolution limit of the XRT.   For these faint and fast moving features this limit is about $10^4$ km.   The observed flow more closely resembles that expected for Petschek-type reconnection than Sweet-Parker reconnection.  A similar conclusion was reached by \citet{vrsnak09} using observations from LASCO.

Future improvements in X-ray and EUV telescopes might eventually make it possible to use flow measurements to differentiate between various reconnection models.  For example, if macro-scale turbulence is present in the current sheet, then it might be possible to see the velocity fluctuations associated with it \citep{mckenzie13}.  At the present time, only a few events exhibit features that can be tracked within flare current sheets \citep{cecere15,schanche16}.

\acknowledgments

   D. B. Seaton's contributions were supported by a grant from the European Union's Seventh Framework Program for Research, Technological Development and Demonstration under grant agreement No. 284461 (Project eHeroes, www.eheroes.eu).  K. K. Reeves' contributions were supported by NSF-SHINE grant AGS-1723425.

\appendix
\section{Applicability of Steady-State Reconnection Equations}

Although the flare model discussed in Section 3 is inherently time dependent, it is possible to use steady-state equations to calculate the reconnection rate if the evolution of the field near the stagnation point is sufficiently slow.  We can determine a necessary condition for this requirement by employing the time-dependent reconnection analysis of \citet{forbes13}.  In that study the time-dependent reconnection rate is determined by a system of three differential equations.  One of these equations is the $y$-component of Faraday's equation averaged across the current sheet, namely:
\begin{equation}
\partial a / \partial t = u_{a} + bV/B_{a} + \eta / a.
\end{equation}
Although the analysis of \citet{forbes13} is incompressible, the same equation holds for the time-dependent compressible system.  Because the reconnection rate is determined by the conditions near the stagnation point, we evaluate Equation~(A1) at $y_{sp}$  to obtain:
\begin{equation}
\partial a_{sp} / \partial t = u_{asp} +  \eta / a_{sp} .
\end{equation}
For time-dependent effects to be completely negligible at the stagnation point, we require that $|\partial a_{sp} / \partial t| \ll |u_{asp}|$, or more precisely we require that $|\partial a_{sp} / \partial t|$ be smaller than   $|u_{asp}|$ by an order of magnitude in the expansion parameter $M_{A}$.  In other words
\begin{equation}
|(\partial a_{sp} / \partial t) / u_{asp}|  <  M_{Asp} .
\end{equation}
If this condition is met, then $u_{asp} \approx - \eta / a_{sp}$ to first order in the expansion.  Therefore, the condition for a steady-state in the vicinity of the stagnation point is
\begin{figure}
\epsscale{0.7}
\plotone{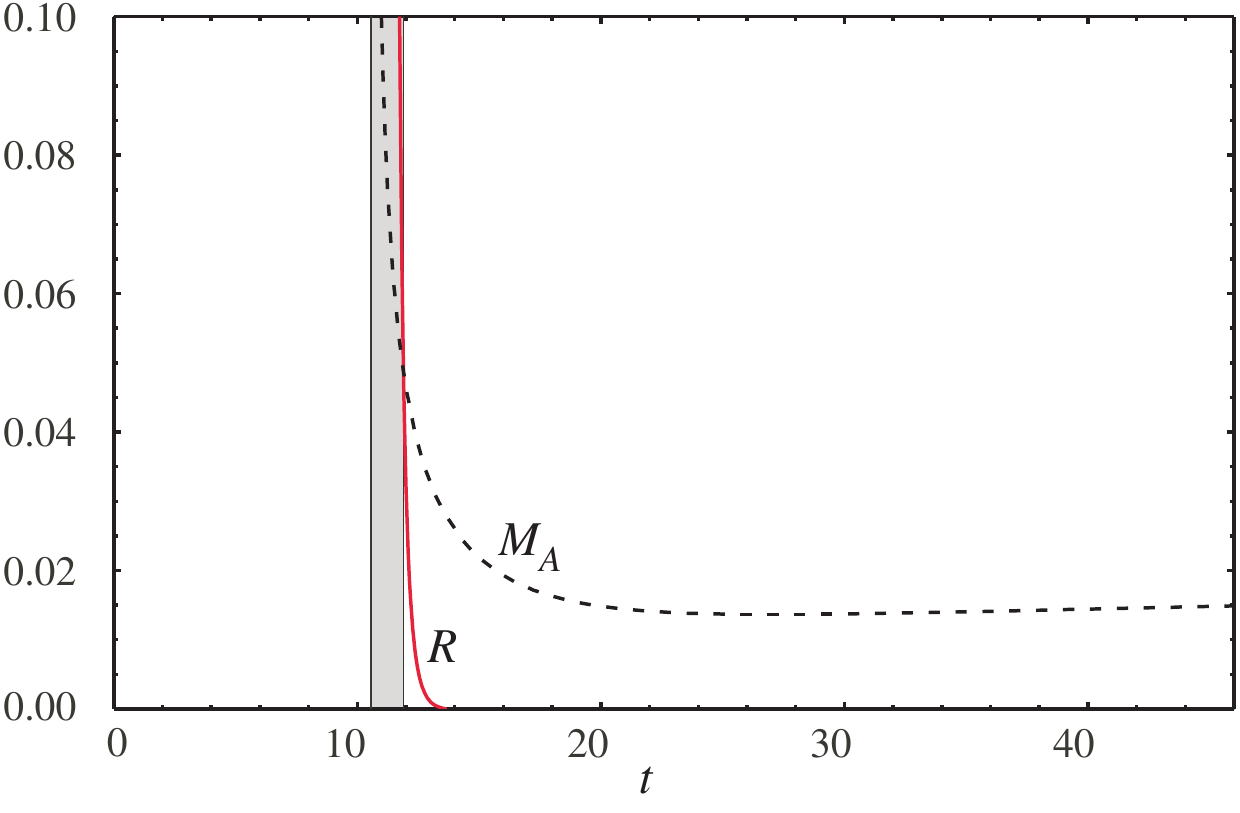}
\caption{Comparison of the time-dependent term, $\partial a_{sp} / \partial t$, to the steady-state term, $u_{asp}$, in Equation~(A2).  When the ratio, $R$, of the time-dependent term to the steady-state term becomes smaller than the inflow Alfv\'{e}n Mach number, $M_{Asp}$, the time-dependent effects in the vicinity of the stagnation point are no longer significant.  The gray shaded region shows the time period after the formation of the current sheet when temporal effects are still important.  This period corresponds to about 1.3 Alfv\'{e}n scale times.}
\end{figure}
\begin{equation}
R  <  M_{Asp} ,
\end{equation}
where the ratio $R$ is defined as
\begin{equation}
R  =   {\partial u_{asp} \over \partial t} {\eta \over u_{asp}^3} = {\partial (M_{Asp} B_{asp}) \over \partial t} \; {\eta \; 4 \pi \rho_{a} \over M_{Asp}^3 B_{asp}^3}.
\end{equation}
Figure 12 shows $R$ and $M_{Asp}$ as functions of time for the case shown in Figure 4.  The shaded region shows the interval from $t = 10.56$ to $t = 11.90$ when the inequality (A4) is not satisfied and time effects are important.  Before 10.56, the current sheet has not yet formed.  After 11.90, time dependent effects are of second order in the expansion parameter $M_{A}$.

Although the lower tip of the current sheet near $p$ can be treated as a quasi-steady structure after $t = 11.90$, the upper tip near, $q$, cannot.  As evident in Figure 4, $q$ moves at about half the speed of the flux rope at $h$.  Depending on the choice of parameters, the speed at which $h$ moves can exceed the ambient Alfv\'{e}n velocity.  If one evaluates Equation~(A1) near $q$ instead of near $p$, the left-hand side is not small.  The reason that it is possible to use the steady-state equation, even though the overall current sheet is not steady, is due to the fact that the nozzle equation, whether steady-state or time-dependent, is an advective equation with the characteristic speed $V$.  Information propagates outwards from the stagnation point, and no information propagates backwards from the tips towards the stagnation point \citep{forbes13}.  If the nozzle equation breaks down because of the onset of instabilities, for example, then the situation is no longer so simple.

\end{document}